\documentclass[aps,prl,reprint,superscriptaddress, fleqn]{revtex4-2}
\usepackage{graphicx}
\usepackage{float}
\usepackage{amsthm} 
\usepackage{amsmath} 
\usepackage{hyperref}
\begin{document}
\title{Implementation and commissioning of an experimental system towards sub-eV axion-like particle searches with 0.1 PW laser at ELI-NP}
\author{Yoshihide Nakamiya}
\email{yoshihide.nakamiya@eli-np.ro}
\affiliation{Extreme Light Infrastructure - Nuclear Physics (ELI-NP), Horia Hulubei National Institute
for R $\&$ D in Physics and Nuclear Engineering (IFIN-HH),
30 Reactorului St., P.O. Box MG-6, 077125 Bucharest-Magurele, Judetul Ilfov, Romania}
\affiliation{Institute for Chemical Research, Kyoto University, Gokasho, Uji, Kyoto 611-0011, Japan}

\author{Kensuke Homma}
\affiliation{Advanced science and engineering, Hiroshima University, 1-3-1 Kagamiyama, 
Higashi-hiroshima, Hiroshima 739-8526, Japan}

\author{Madalin-Mihai Rosu}
\affiliation{Extreme Light Infrastructure - Nuclear Physics (ELI-NP), Horia Hulubei National Institute
for R $\&$ D in Physics and Nuclear Engineering (IFIN-HH),
30 Reactorului St., P.O. Box MG-6, 077125 Bucharest-Magurele, Judetul Ilfov, Romania}

\author{Liviu Neagu}
\affiliation{Extreme Light Infrastructure - Nuclear Physics (ELI-NP), Horia Hulubei National Institute
for R $\&$ D in Physics and Nuclear Engineering (IFIN-HH),
30 Reactorului St., P.O. Box MG-6, 077125 Bucharest-Magurele, Judetul Ilfov, Romania}
\affiliation{CETAL, National Institute for Laser, Plasma and Radiation Physics, 
409 Atomistilor, P.O. Box MG-36, 077125 Bucharest-Magurele, Judetul Ilfov,
Romania}

\author{Mihai Cuciuc}
\author{Vanessa Rozelle Maria Rodrigues}
\author{Stefan Ataman}
\author{Catalin Chiochiu}
\affiliation{Extreme Light Infrastructure - Nuclear Physics (ELI-NP), Horia Hulubei National Institute
for R $\&$ D in Physics and Nuclear Engineering (IFIN-HH),
30 Reactorului St., P.O. Box MG-6, 077125 Bucharest-Magurele, Judetul Ilfov, Romania}

\author{Georgiana Giubega}
\affiliation{Extreme Light Infrastructure - Nuclear Physics (ELI-NP), Horia Hulubei National Institute
for R $\&$ D in Physics and Nuclear Engineering (IFIN-HH),
30 Reactorului St., P.O. Box MG-6, 077125 Bucharest-Magurele, Judetul Ilfov, Romania}
\affiliation{CETAL, National Institute for Laser, Plasma and Radiation Physics, 
409 Atomistilor, P.O. Box MG-36, 077125 Bucharest-Magurele, Judetul Ilfov,
Romania}

\author{Kyle Juedes}
\author{Jonathan Tamlyn}
\author{Stefan Victor Tazlauanu}
\affiliation{Extreme Light Infrastructure - Nuclear Physics (ELI-NP), Horia Hulubei National Institute
for R $\&$ D in Physics and Nuclear Engineering (IFIN-HH),
30 Reactorului St., P.O. Box MG-6, 077125 Bucharest-Magurele, Judetul Ilfov, Romania}

\author{Yuri Kirita}
\affiliation{Advanced science and engineering, Hiroshima University, 1-3-1 Kagamiyama, 
Higashi-hiroshima, Hiroshima 739-8526, Japan}

\author{Akihide Nobuhiro}
\affiliation{Advanced science and engineering, Hiroshima University, 1-3-1 Kagamiyama, 
Higashi-hiroshima, Hiroshima 739-8526, Japan}

\author{Masaki Hashida}
\affiliation{Institute for Chemical Research, Kyoto University, Gokasho, Uji, Kyoto 611-0011, Japan}
\affiliation{Graduate School of Science, Kyoto University, Kitashirakawa Oiwake-cho, Sakyo-ku, Kyoto 606-8502, Japan}
\affiliation{Research Institute of Science and Technology, Tokai University, 4-1-1 Kitakaname, Hiratsuka, Kanagawa 259-1292, Japan}

\author{Shunsuke Inoue}
\affiliation{Institute for Chemical Research, Kyoto University, Gokasho, Uji, Kyoto 611-0011, Japan}
\affiliation{Graduate School of Science, Kyoto University, Kitashirakawa Oiwake-cho, Sakyo-ku, Kyoto 606-8502, Japan}

\author{Shuji Sakabe}
\affiliation{Institute for Chemical Research, Kyoto University, Gokasho, Uji, Kyoto 611-0011, Japan}
\affiliation{Graduate School of Science, Kyoto University, Kitashirakawa Oiwake-cho, Sakyo-ku, Kyoto 606-8502, Japan}

\author{Vicentiu Iancu}
\affiliation{Extreme Light Infrastructure - Nuclear Physics (ELI-NP), Horia Hulubei National Institute
for R $\&$ D in Physics and Nuclear Engineering (IFIN-HH),
30 Reactorului St., P.O. Box MG-6, 077125 Bucharest-Magurele, Judetul Ilfov, Romania}
\affiliation{Faculty of Physics, University of Bucharest, 405 Atomistilor Street, 077125 Magurele, Romania}

\author{Dan-Gheorghita Matei}
\affiliation{Extreme Light Infrastructure - Nuclear Physics (ELI-NP), Horia Hulubei National Institute
for R $\&$ D in Physics and Nuclear Engineering (IFIN-HH),
30 Reactorului St., P.O. Box MG-6, 077125 Bucharest-Magurele, Judetul Ilfov, Romania}

\author{Mihai Risca}
\affiliation{Extreme Light Infrastructure - Nuclear Physics (ELI-NP), Horia Hulubei National Institute
for R $\&$ D in Physics and Nuclear Engineering (IFIN-HH),
30 Reactorului St., P.O. Box MG-6, 077125 Bucharest-Magurele, Judetul Ilfov, Romania}

\author{Anda-Maria Talposi}
\affiliation{Extreme Light Infrastructure - Nuclear Physics (ELI-NP), Horia Hulubei National Institute
for R $\&$ D in Physics and Nuclear Engineering (IFIN-HH),
30 Reactorului St., P.O. Box MG-6, 077125 Bucharest-Magurele, Judetul Ilfov, Romania}
\affiliation{Faculty of Physics, University of Bucharest, 405 Atomistilor Street, 077125 Magurele, Romania}

\author{Ovidiu Tesileanu}
\affiliation{Extreme Light Infrastructure - Nuclear Physics (ELI-NP), Horia Hulubei National Institute
for R $\&$ D in Physics and Nuclear Engineering (IFIN-HH),
30 Reactorului St., P.O. Box MG-6, 077125 Bucharest-Magurele, Judetul Ilfov, Romania}

\author{Kazuo A. Tanaka}
\affiliation{Extreme Light Infrastructure - Nuclear Physics (ELI-NP), Horia Hulubei National Institute
for R $\&$ D in Physics and Nuclear Engineering (IFIN-HH),
30 Reactorului St., P.O. Box MG-6, 077125 Bucharest-Magurele, Judetul Ilfov, Romania}
\affiliation{Institute of Laser Engineering, Osaka University, Yamada-oka 2-6, Suita, Osaka 565-0871 Japan}
\affiliation{National University of Science and Technology POLITEHNICA Bucharest, 313 Splaiul Independentei, 060042 Bucharest, Romania}

\author{Calin Alexandru Ur}
\affiliation{Extreme Light Infrastructure - Nuclear Physics (ELI-NP), Horia Hulubei National Institute
for R $\&$ D in Physics and Nuclear Engineering (IFIN-HH),
30 Reactorului St., P.O. Box MG-6, 077125 Bucharest-Magurele, Judetul Ilfov, Romania}
\collaboration{The SAPPHIRES collaboration and ELI-NP operation team}

\begin{abstract}%
We have developed and commissioned an experimental system at ELI-NP towards searches for axion-like particles (ALPs) in the worldwide 10~PW-class laser facility. 
The search principle is based on the Four-Wave Mixing (FWM) process 
at a focal region of coaxially combined two laser beams.
The subsystems to control vacuum pressure, area size, spatiotemporal overlap and trigger-event pattern,  
are integrated into the experimental area for 0.1 PW laser output at ELI-NP. 
The integrated system is dedicated to identifying the possible background sources originated from the residual atoms and the optical elements. 
The performance and functionality of the subsystems were validated 
through the evaluations of laser characteristics, their stability and the FWM signal detections. 
Furthermore commissioning results for the background studies were demonstrated
with 20 mJ-level laser pulses at the vacuum pressure of $1.3 \times 10^{-7}$ mbar.

In conclusion, the integrated experimental system is fully functional as designed 
and provides a suitable platform for the background studies towards the ALP searches, enabling 
a stepwise scale-up of the laser pulse energies from 20 mJ to the maximum energy of 2.5 J in the 0.1 PW experimental area. 
\end{abstract}

\maketitle

\section{Introduction}
Dark Matter (DM) remains one of the most important unsolved problems in modern physics,
in particular, cold dark matter (CDM) is favored among the many proposed candidates because its non-relativistic nature explains both the observed dark-matter abundance 
and the formation of large-scale structures in the universe \cite{Cold_DM1}. 
Among the CDM candidates, axion is considered to play an important role in resolving the strong CP problem \cite{Strong_CP1,Strong_CP2} via Peccei-Quinn symmetry breaking  \cite{PQ_Symmetry}. Axion and, more generically, some of axion-like particles (ALPs) are regarded as pseudo-Nambu-Goldstone bosons which appear with small masses. Depending on cosmological conditions, axion and ALPs can occupy a viable mass range extending from the $\mu$eV scale up to the sub-eV region. This motivates direct laboratory searches for axion and ALPs in the meV-to-sub-eV mass range.

\begin{figure}[h]
 \centering
 \includegraphics[width=\columnwidth]{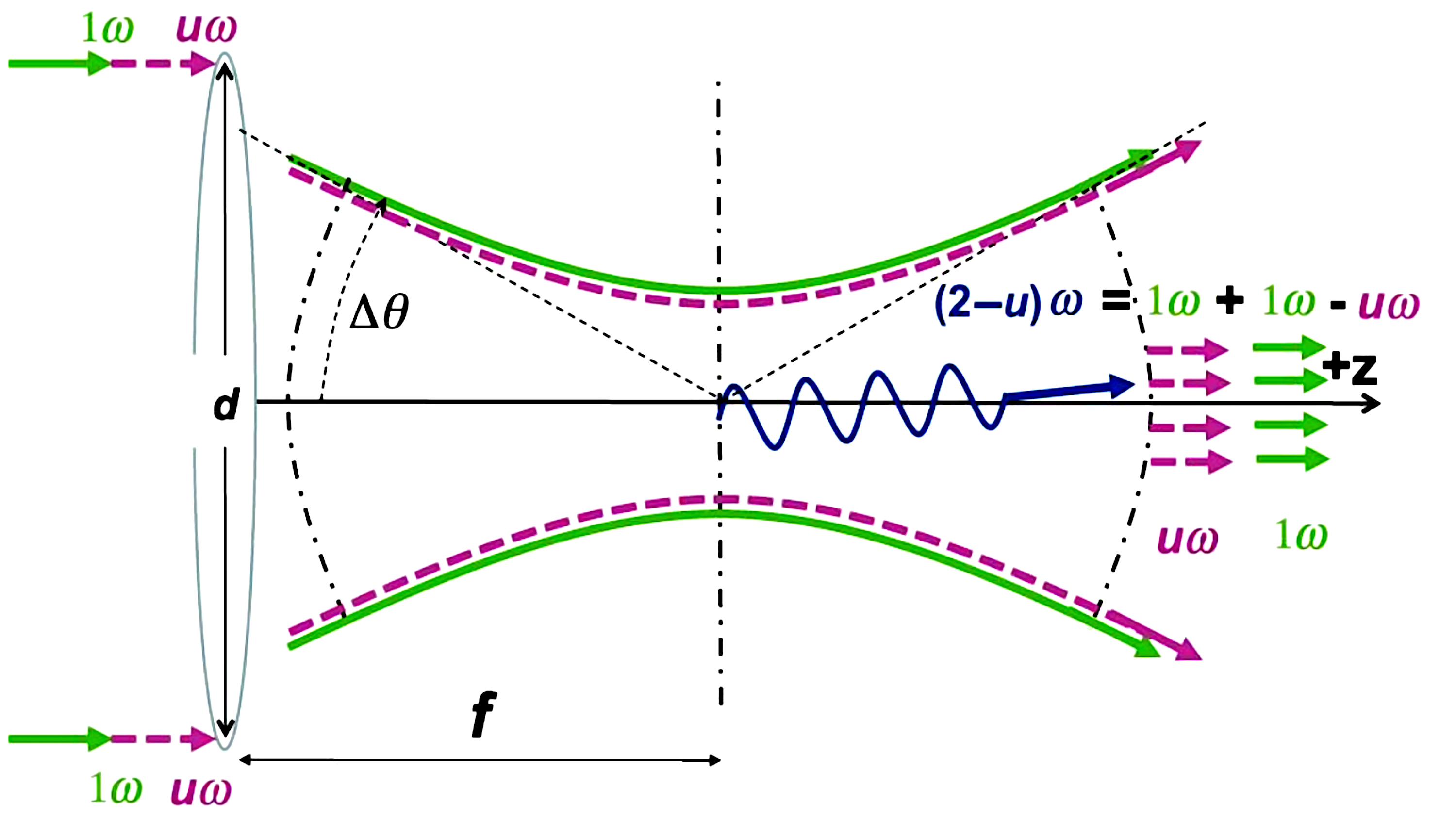}\label{fig_FWM}
 \caption{Four-wave mixing process for two focused lasers with different wavelength. 
 Solid green arrow shows a "creation" laser.
 Arbitrary two photons in single "creation" laser beam can collide with center-of-mass energy 
 up to $2\omega\sin{\Delta \theta}$. 
 Dotted magenta arrow shows an "inducing" laser for the stimulated decay into final-state photons (Blue). 
 Reproduced from Ref. \cite{PTEP2015_10uJ}.}
 \end{figure}

Laser-driven searches have an excellent potential in discovering ALPs in the sub-eV mass range against the extremely small coupling constant associated with ALP productions \cite{KH_DH_TT,KH,YF_KH}. 
In particular, high-power short-pulse lasers may increase the chance to produce ALP resonant states via photon-photon interactions,
because the huge number of photons can be focused into a small interaction volume comparable to the wavelength scale of optical laser photons. 
In addition to photon-photon collisions within a single focused laser beam ("creation" laser), the other laser beam ("inducing" laser) at a different wavelength from that of the creation laser can simultaneously induce the decay of the produced resonance states at around the production point. The entire two-body scattering process is thus interpreted as a stimulated resonant photon-photon scattering process~\cite{KH_YK}. 
The inducing laser plays a critical role in stimulating the decays and determining the wavelength 
of the final-state decay photons as illustrated in Fig. \ref{fig_FWM}. 
The mechanism is also referred to as four-wave mixing (FWM) process in vacuum in the laser terminology. 

The probability of observing a photon-photon interaction through the ALP production has been enhanced 
by the remarkable development of short-pulse high-intensity laser technology over the past several decades 
\cite{PWLaser1,PWLaser2}, since the production yield has a quadratic dependence on the intensity of the creation laser 
and a linear dependence on the intensity of the inducing laser in the FWM process. 
In particular, the ALPs search with PW-class high-power laser systems has the promising potential to reach 
a theoretical prediction for the QCD axions in the sub-eV mass range \cite{EPJ2023}.
The four-wave mixing approach with two laser pulses focused in space and time provides a unique opportunity 
to explore the interaction for both on-shell and off-shell ALP production and offers a complementary study 
in comparison to the astrophysical experiments such as solar axion experiments \cite{CAST1,CAST2,CAST3,CAST4}, 
“Light Shining through Wall (LSW)” experiments \cite{LSW1,LSW2}, and others. 

We have updated the exclusion limits of the ALPs by the FWM approach 
with the laser pulse energies from $\mu$J to mJ using a 10 TW Ti:Sa laser system 
at Institute of Chemical Research in Kyoto university \cite{PTEP2015_10uJ, JHEP2021_SAPPHIRES_100uJ,JHEP2022_SAPPHIRES_1mJ,JHEP2025_SAPPHIRES_10mJ}. 
As reported in our latest research \cite{JHEP2025_SAPPHIRES_10mJ}, we reached 
the exclusion region at the ALPs-photon coupling 
$ g/M = 5.45 \times 10^{-7}$ GeV $^{-1}$ and at the ALPs mass $m$ = 0.15 eV. 
Increasing laser intensity is a key direction to access the unexplored lower-coupling limit of the ALP production. 
In particular, upgrading to a PW-class laser system is a great step towards reaching the theoretical QCD prediction 
and beyond the predicted region.  
However, the upgrade from a TW laser system to a PW one requires a paradigm shift from "table-top" to "facility-scale" experiments,
in terms of handling the backgrounds in the presence of high-intensity laser fields and controlling 
systematic errors of such a large-scale experimental system.
In this article, we classify key design issues for the upgrade towards a PW-class experiment in terms 
of the background studies, demonstrate the performance of the experimental setups integrated into the 0.1 PW experimental area 
at ELI-NP, and present the results of FWM observation at the 20 mJ-level pulse energies 
during a commissioning experiment.

\section{Key issues of experimental system for the background studies} 
Continuous developments of laser intensity enable us to approach lower coupling limits 
below the predicted range of the QCD axion.
However, increasing intensity can also increase the background level, which has remained unobserved 
in the previous experiments with relatively low-intensity laser pulses, 
even when sufficient statistics were accumulated for the null hypothesis test 
to address the exclusion limits of the ALP production. 
Furthermore, upgrading the laser system to a PW-class power requires the facility-scale infrastructure.
Under such situations, handling experimental systems gets more complicated, in particular, 
for relatively high-statistics experiments such as the ALPs search.
An important challenge lies in fine control of the initial conditions for the FWM interaction, 
such as spatiotemporal overlap between two lasers.  
In this section, we discuss the key issues for the development of the experimental system with 
clarifying of the possible background sources.

\begin{figure}[!h]
\centering
\includegraphics[width=\columnwidth]{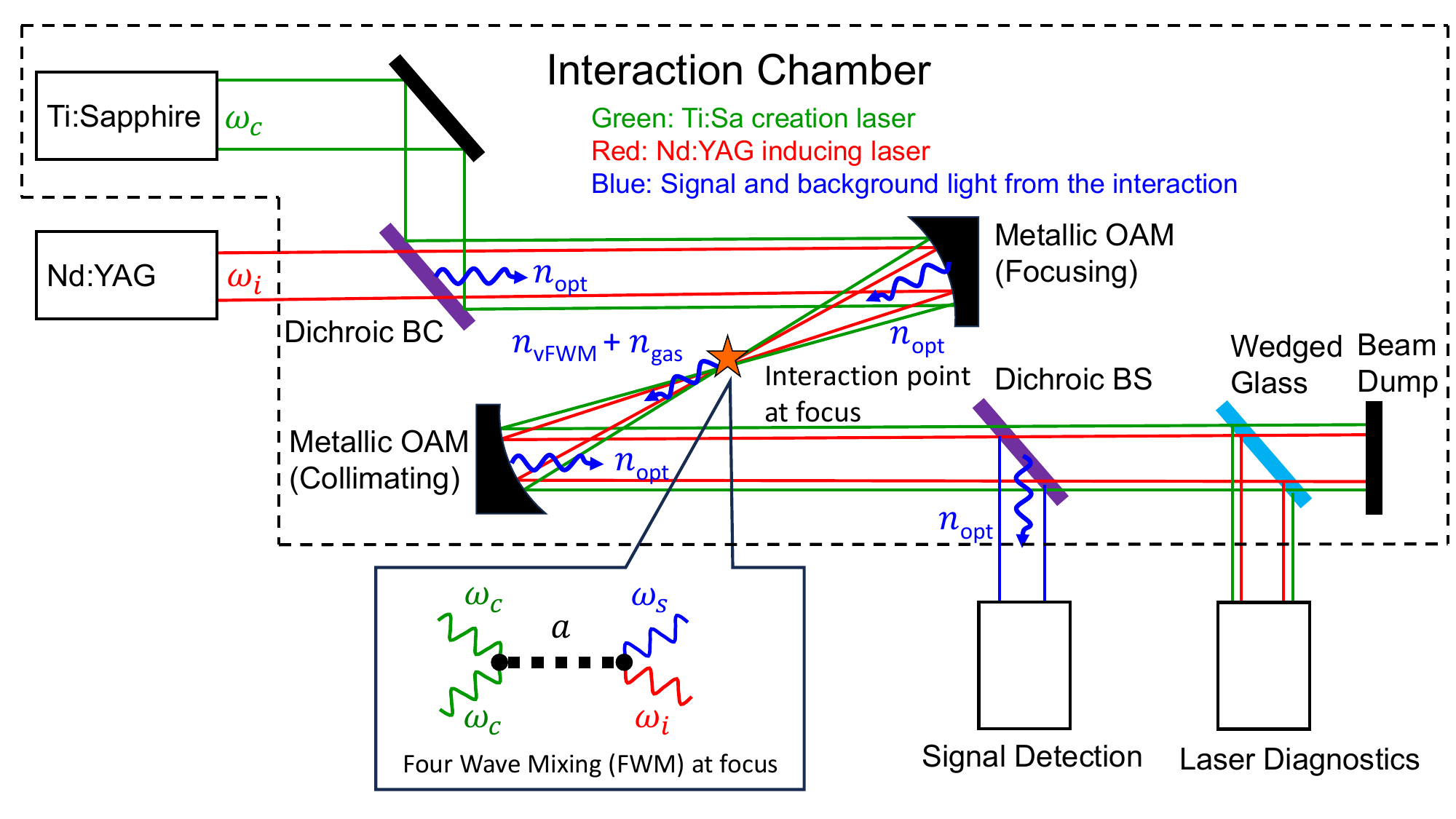}
\caption{A conceptual setup for a laser-driven ALPs search by the FWM approach.}
\label{fig_conceptual_E4}
\end{figure}

The conceptual setup of laser-driven ALPs search via FWM approach is shown in Fig. \ref{fig_conceptual_E4}.
In this scheme, Ti:Sa creation laser ($\lambda_c$ = 800 nm) and Nd:YAG inducing laser ($\lambda_i$ = 1064 nm) 
are simultaneously focused by using a single set of focusing optics 
in the quasi-parallel photon-photon collision geometry, which is characterized by the focusing angle 
(i.e., F-number). 
The focusing parameter and laser wavelength in our condition target low-mass ALP production below 1 eV. 
We typically employ the metallic off-axis parabolic mirrors as focusing and collimation optics 
rather than refractive lenses to prevent unwanted nonlinear effects and subsequent generation 
of the background photons in the laser propagation through the glass substrate.  
In the FWM process, the wavelength of the signal photon is expected to be $\lambda_{s} = 641$ nm, as determined 
by the energy conservation such that 
\begin{equation}
    \lambda_{s}=\frac{\lambda_c\lambda_i}{2\lambda_i-\lambda_c}.
\end{equation}  
The signal photons are isolated from residual laser beams through wavelength selection 
by using the dichroic mirrors and the band pass filter stacks prior to the signal detection. 
This spectral filtering also works to reject the other scattered background light existing in the experimental area. 

The possible processes which can generate the photons at the same wavelength as the FWM signal are categorized 
as follows (See also Table \ref{table_BG_process}).
The first process involves the vacuum-origin photons ($n_{vFWM}$), which are the desired signal originating from the ALP production. 
These photons are produced at the focal spot where the high-intensity laser field is present. 

\begin{figure}[!h]
\centering
\includegraphics[scale=0.27]{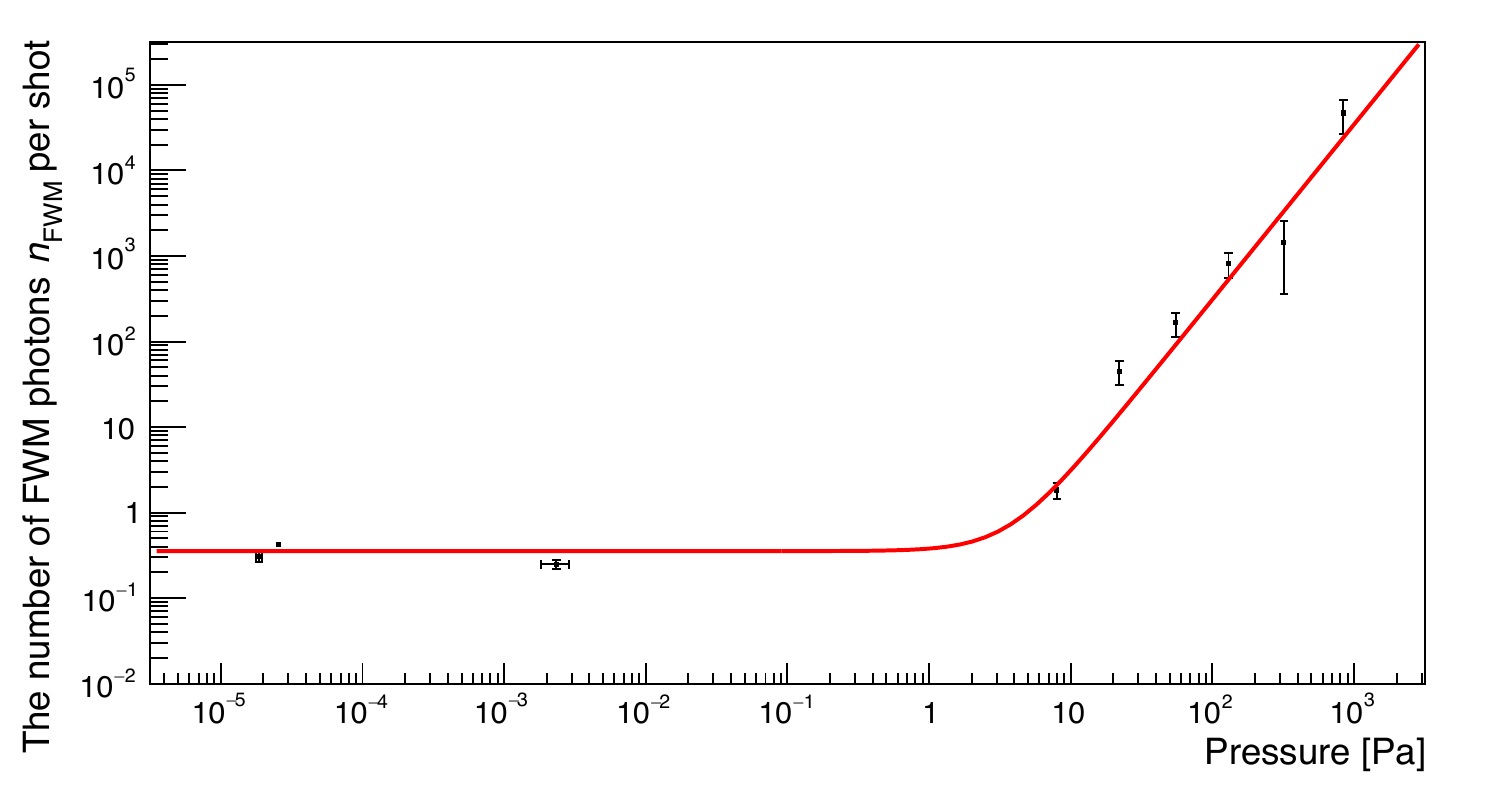}
\includegraphics[scale=0.26]{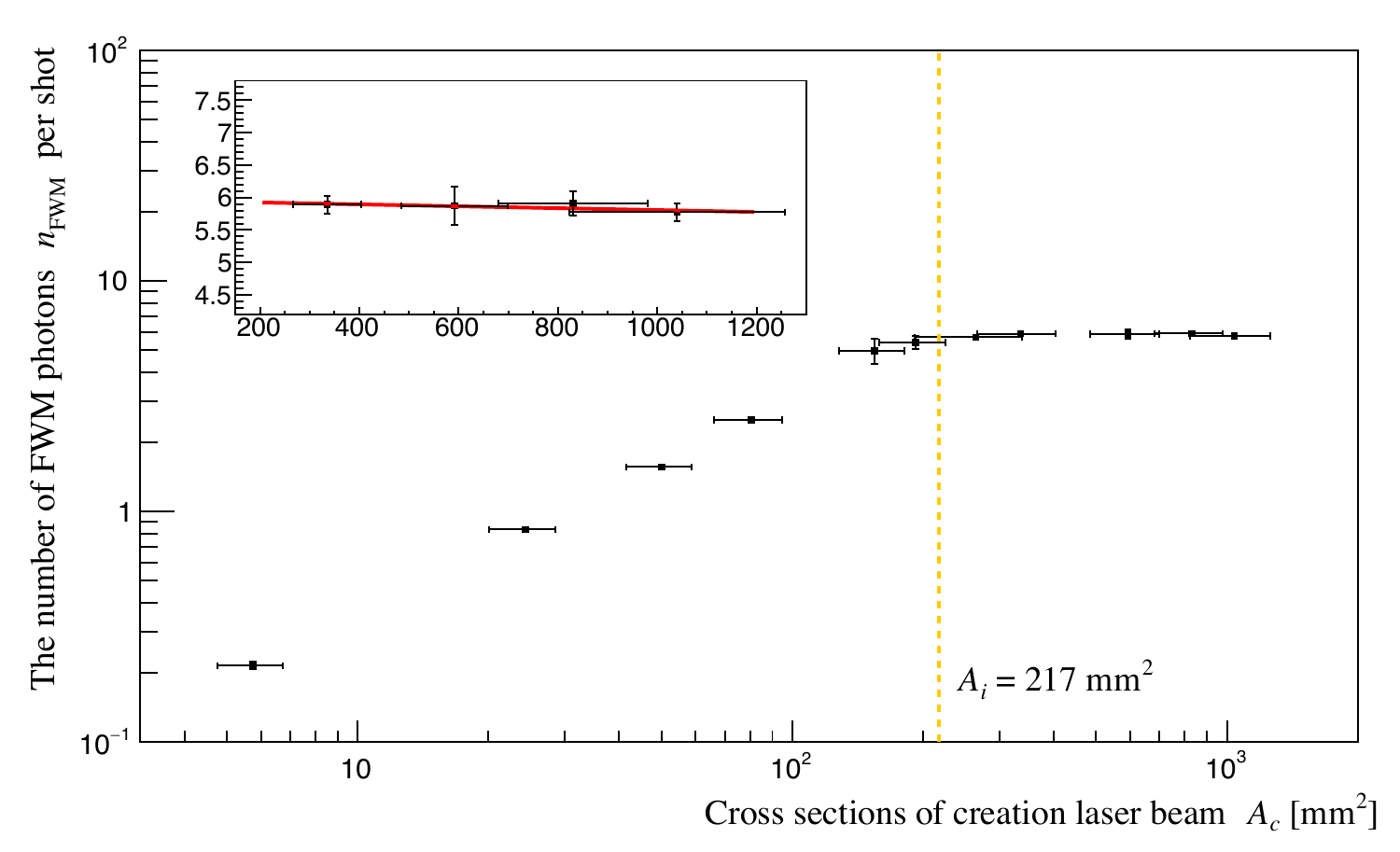}
\caption{(Top) Pressure dependence to evaluate atomic four-wave mixing generation. 
(Bottom) Area-size dependence to evaluate optical four-wave mixing generation. 
Reproduced from Ref. \cite{JHEP2022_SAPPHIRES_1mJ}.}
\label{fig_bg_species}
\end{figure}

The second process involves the gas-origin photons ($n_{gas}$) resulting from 
the interaction between the intense laser pulses and the residual molecules and atoms in the vacuum chamber. 
Since the gas-origin photon yield nonlinearly depends on the laser intensity, they are also primarily produced 
at the focal spot. The gas-origin photons are further classified into atomic FWM photons ($n_{aFWM}$) 
and the plasma-induced photons ($n^{gas}_{plasma}$) by laser-gas interaction.
The plasma-induced photons are mainly produced by Ti:Sa creation laser rather 
than Nd:YAG inducing laser,  
since the intensity of Ti:Sa laser is several orders of magnitude higher than that of Nd:YAG laser. 
Thus, the contribution of this process can be evaluated and subtracted by comparing different triggered events:
the events where both lasers are active and those where only one of them is active. 
A common aspect of the gas-origin photons is, whatever the processes are, the photons follow
 a power-law scaling as a function of operational vacuum pressure $P^{\alpha}$, for instance, 
 $\alpha = 2.05 \pm 0.09$ \cite{JHEP2022_SAPPHIRES_1mJ}.
The top plot of Fig. \ref{fig_bg_species} shows the pressure dependence of the FWM generation observed 
in our past experiment with mJ laser pulse energy \cite{JHEP2022_SAPPHIRES_1mJ}. 
A power-law scaling is clearly observed as the pressure-dependent FWM generation in the high pressure range above 1 Pa ($10^{-2}$ mbar). 

In contrast, the pressure-independent photon generation becomes relatively significant 
under a high-vacuum condition such that the gas-origin background generation is suppressed, 
for example, below 10$^{-2}$ Pa in the top plot of Fig. \ref{fig_bg_species}. 
Candidate sources of pressure-independent photons are the optical elements ($n_{opt}$), 
or potentially ALPs.
Such optics-origin photons are also classified into the optical FWM photons ($n_{oFWM}$) 
and the plasma-induced photons on the mirror surface ($n^{opt}_{plasma}$). 
The generation of optics-origin photons can occur at any mirror placed along the common path 
after Ti:Sa creation laser and Nd:YAG inducing laser are combined. 
In particular, the dichroic beam splitter ("Dichroic BS" shown in Fig. \ref{fig_conceptual_E4}) 
for the isolation of the FWM signal from residual laser pulses is expected to be the primary contributor, 
since the dichroic mirror has, by definition, finite susceptibilities to enable a parametric conversion 
such as the FWM process when the two lasers transmit through the dielectric layers and the glass substrates. 
The other metallic focusing and collimating mirrors are considered to have a smaller impact on FWM generation 
from this viewpoint due to the non-dielectric nature of their coating materials. 
Furthermore, the spatial overlap of the two lasers during the reflection is limited to the skin depth, 
which is on the order of 100 nm at most. 
This could suppress the interaction rate for FWM generation, even if it takes place within the coating layers. 
However, we cannot exclude the scenario of other parametric processes arising from the metallic surfaces 
via laser-induced plasma on the metallic surface when the mirror is irradiated by high-power laser fields. 
Additionally, a secondary FWM generation process may also take place from the energy deposition of laser pulses 
because of the intrinsic absorption of the metal. 
These scenarios should be taken into consideration, although their contributions remain unclear 
at the current laser power.
Regarding the plasma-induced photons ($n^{opt}_{plasma}$), their spectrum is expected to 
be supercontinuum (broadband). Thus, only a small fraction of the generation would contaminate the signal wavelength range  
in the observation.
These background photons ($n_{oFWM}$ and $n^{opt}_{plasma}$) are also distinguishable 
by trigger-event selections in the same way explained in the paragraph on the gas-origin background photons 
($n^{gas}_{plasma}$). 

Another specific case is the background photon emission around a wavelength of 650 nm 
from the fused silica substrate ($n_{FS}$). It is caused by multi-photon absorption, excitation and subsequent de-excitation 
through a defect level, attributed to the formation of a non-bridging oxygen hole center (NBOHC)\cite{FS_Defect}. 

\begin{figure}[!h]
\centering
\includegraphics[width=\columnwidth]{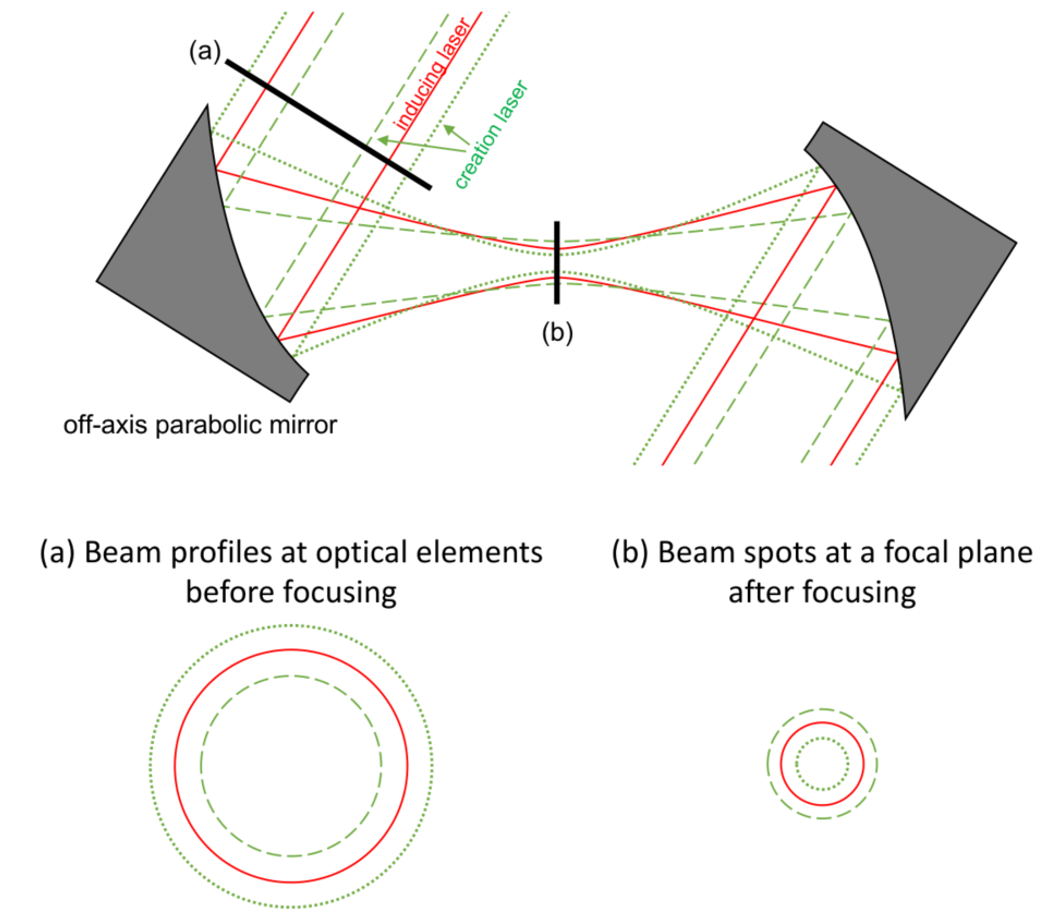}
\caption{Spatial overlap between Ti:Sa creation laser (green) and Nd:YAG inducing laser (red) 
in the laser propagation through focusing and collimating by two OPMs. 
Reproduced from Ref.  \cite{JHEP2022_SAPPHIRES_1mJ}.}
\label{fig_areasize}
\end{figure}

Further evaluation of the pressure-independent background photons from the optical elements can be studied 
by an area-size scan as shown in the bottom plot of Fig. \ref{fig_bg_species}. 
The area-size scan is designed to change spatial overlap by controlling the beam diameter 
of Ti:Sa laser. The beam cross-section of the two lasers is illustrated with their spatial overlap before and after focusing 
in Fig. \ref{fig_areasize}. 
The red circles represent the edge of Nd:YAG laser beam profile before focusing (a) and at the focal spot (b).
The green dotted line indicates the edge of Ti:Sa laser beam  at $A_c > A_i$, and the green dashed line indicates the beam edge at $A_c < A_i$,  
where $A$ denotes the area size of the laser and the subscripts $c$ and $i$ show creation (Ti:Sa) and inducing (Nd:YAG), respectively. 
Assuming the gas-origin backgrounds are sufficiently suppressed at low vacuum pressures, 
the optics-origin background yield is expected to increase monotonically with the area size $A_c$ 
for $A_c \leq A_i$ and reach a plateau for $A_c > A_i$. 
In contrast, the ALPs signal yield should continue to increase even at $A_c > A_i$.
This is because increasing the laser diameter reduces the focal-spot size and 
the effective energy density in the focal-spot area is enhanced. 
This enhancement outweighs the reduction of the overlap area size with Nd:YAG laser 
in terms of the ALP production. 

In summary, the observable photons in a laser-driven ALPs search are categorized by the equations below. 
\begin{eqnarray}
    n_{tot} &=& n_{vFWM} + n_{gas} + n_{opt}, \label{eq1} \\ 
    n_{gas} &=& n_{aFWM} + n^{gas}_{plasma}, \label{eq2} \\
    n_{opt} &=& n_{oFWM} + n^{opt}_{plasma} + n_{FS}, \label{eq3}
\end{eqnarray} 
where $n_{tot}$ is the total number of photons including the signal process and all the background processes, 
which favor the generation of the photons around 641 nm.

\begin{table*}[t]
 \caption{Signal and background photons around $\lambda_s=$ 641 nm. BC, BS, OPM 
 are abbreviations for dichroic Beam Combiner, dichroic Beam Splitter and metallic Off-axis Parabolic Mirror, 
 respectively.}%
 \label{table_BG_process}
 \centering
 \begin{tabular}{|l|l|l|l|l|}
 \hline
 Process  & Source & Event Class & Scaling  \\ 
 \hline
 \hline
 ALPs signal ($n_{vFWM}$)         & ALPs at focus & Ti:Sa $\cap$ Nd:YAG &  Area size \\
 \hline
 Atomic FWM ($n_{aFWM}$)          & Molecules at focus & Ti:Sa $\cap$ Nd:YAG &  Pressure $\&$ Area size \\
 \hline
 Atomic plasma ($n^{gas}_{plasma}$)  & Molecules at focus & Ti:Sa & Pressure $\&$ Area size \\
 \hline
 Optics FWM ($n_{oFWM}$)          & BC, BS, OPM  & Ti:Sa $\cap$ Nd:YAG  & Area size \\
 \hline
 Optics plasma($n^{opt}_{plasma}$)  & BC, BS, OPM  & Ti:Sa & Area size \\
 \hline
 Defect level in glass ($n_{FS}$)   & BS  & Ti:Sa & Area size \\
 \hline

\end{tabular}
 \end{table*}

To investigate the scaling properties of the background processes with multiple experimental parameters 
such as vacuum pressure and area size. We require datasets exceeding $10^{5}$ laser shots, 
which is a much higher number than in other nominal single-shot basis experiments with 
PW-class high-power lasers. 
Thus, careful handling of the experimental system and the initial conditions for the laser interaction are crucial 
for such situations. For example, the spatiotemporal overlap of two lasers should be precisely controlled 
at the focal point. The shot-by-shot yield of the vacuum-origin FWM generation (i.e., ALPs signal) is parameterized by
\begin{eqnarray}
  \mathcal{Y}_{c+i} = \frac{1}{4}N_{c}^{2}N_{i} \mathcal{D}_{I}\bar{\Sigma}_{I}, \label{eq:FEMyeild}
\end{eqnarray} 
where $N_{c}$ and $N_{i}$ represent the average number of photons in the creation and inducing lasers, respectively.
The product $\mathcal{D}_{I}$$\bar{\Sigma}$ denotes the efficiency of the photon collision. 
$\mathcal{D}_{I}$ is the spatiotemporal overlap factor between the two lasers and  $\bar{\Sigma}$ is 
the interaction volume rate.
Laser instability can directly impact the spatiotemporal overlap factor $\mathcal{D}_{I}$, 
which is defined by
\begin{equation}
  \mathcal{D}_{I} = \int_{0}^{t'}dt \int_{-\infty}^{\infty} d\bm{r}\rho_{c}\left(\bm{r},t \right)\rho_{c}\left(\bm{r},t \right)\rho_{i}\left(\bm{r},t \right)V_{i}, \label{eq:space-time_overlap}
\end{equation} 
Assuming a spatiotemporal Gaussian beam for the two lasers,
\begin{equation}
  \rho_{k}\left(\bm{r},t \right) =  A_c\exp \left(-2\frac{x^2+y^2}{\omega^{2}_{k}\left(ct\right)}\right)
  \exp\left(-2\frac{\left(z-ct\right)^{2}}{c\tau_{k}}\right), \label{space_c} 
\end{equation}
where the subscript $k$ is $c$ to represent Ti:Sa creation laser or $i$ to represent Nd:YAG inducing laser. 
Two important design issues can be raised regarding the spatiotemporal overlap factor. 
The first issue is how to overlap in space at the focus to keep peak-to-peak interactions 
despite the pointing instability inherent in the high-power laser system.
We designed the beam diameter of Nd:YAG laser to be three times larger than that of Ti:Sa creation laser 
at the focus. This design effectively suppresses the shot-by-shot systematic uncertainty 
of the spatial overlap factor.
As a trade-off, the design results in creating non-interacting photons outside of the overlapped area. 
Nevertheless, minimizing this systematic error is prioritized because 
the loss for the ineraction can be compensated by the high energy output from the 3J high-power Nd:YAG laser. 
The second issue is how to synchronize the timing between the two lasers. 
Since the pulse duration of Ti:Sa laser ($\sim$ 25 fs) is negligibly short 
compared to that of Nd:YAG laser ($\sim$ 10 ns), timing synchronization can be managed   
electronically,  as long as the low-jitter timing control devices are implemented. 

We designed and implemented the experimental system at the 0.1 PW experimental area at ELI-NP according 
to the requirements discussed in this section. The details of the implemented setup are presented 
in the following sections. 

\section{Implementation and performance of the experimental system}
\subsection{High-power laser system}
\begin{figure}[H]
\centering
\includegraphics[width=\columnwidth]{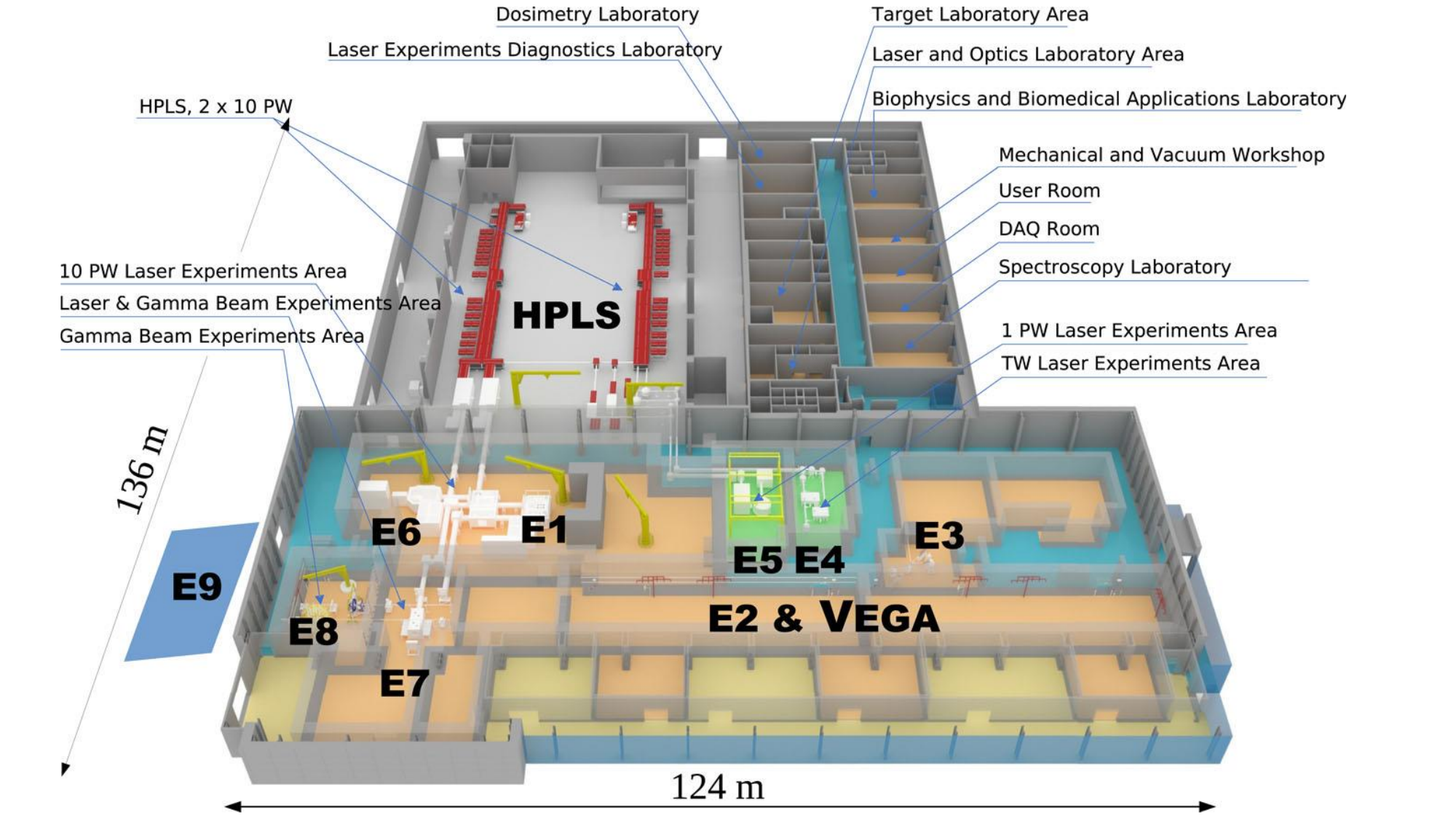}
\caption{Layout of ELI-NP facility. Reproduced from Ref. \cite{KAT_ELI_NP}.}
\label{fig_ELI-NP}
\end{figure}

The ELI-NP facility has the dual-arm High-Power Laser System (HPLS), which is capable of generating 
femtosecond laser pulses at peak power levels of 0.1 PW, 1 PW and 10 PW. 
The HPLS is a hybrid amplification system composed of Chirped Pulse Amplification (CPA) 
\cite{CPA} and Optical Parametric Chirped Pulse Amplification (OPCPA) \cite{OPCPA}. 
The CPA system employs multiple amplification stages with Ti:Sapphire crystal for sub-ns-stretched broadband laser pulse 
to reach 250 J laser energy. The OPCPA system utilizes a parametric amplification 
for broadband gain and the suppression of the amplified spontaneous emission (ASE) 
prior to seeding the main CPA stages. 
The amplified laser pulses are compressed to a pulse duration of around 25 fs 
using diffraction gratings at the exit of the HPLS.
The operational performance of the HPLS at ELI-NP was reported in the publications \cite{HPLE2022_ELI-NP,HPLE2024_ELI-NP}.
Figure \ref{fig_ELI-NP} shows the layout of the ELI-NP facility. 
Two ultra-short high-power laser pulses are delivered from the HPLS to several experimental areas (E1 to E9) 
located in the lower section of the facility at a distance of approximately 100 m from the HPLS front-ened.  
The experimental system for the ALPs search is implemented 
at the E4 experimental area dedicated to the 0.1 PW laser output.

\begin{figure}[H]
\centering
\includegraphics[width=\columnwidth]{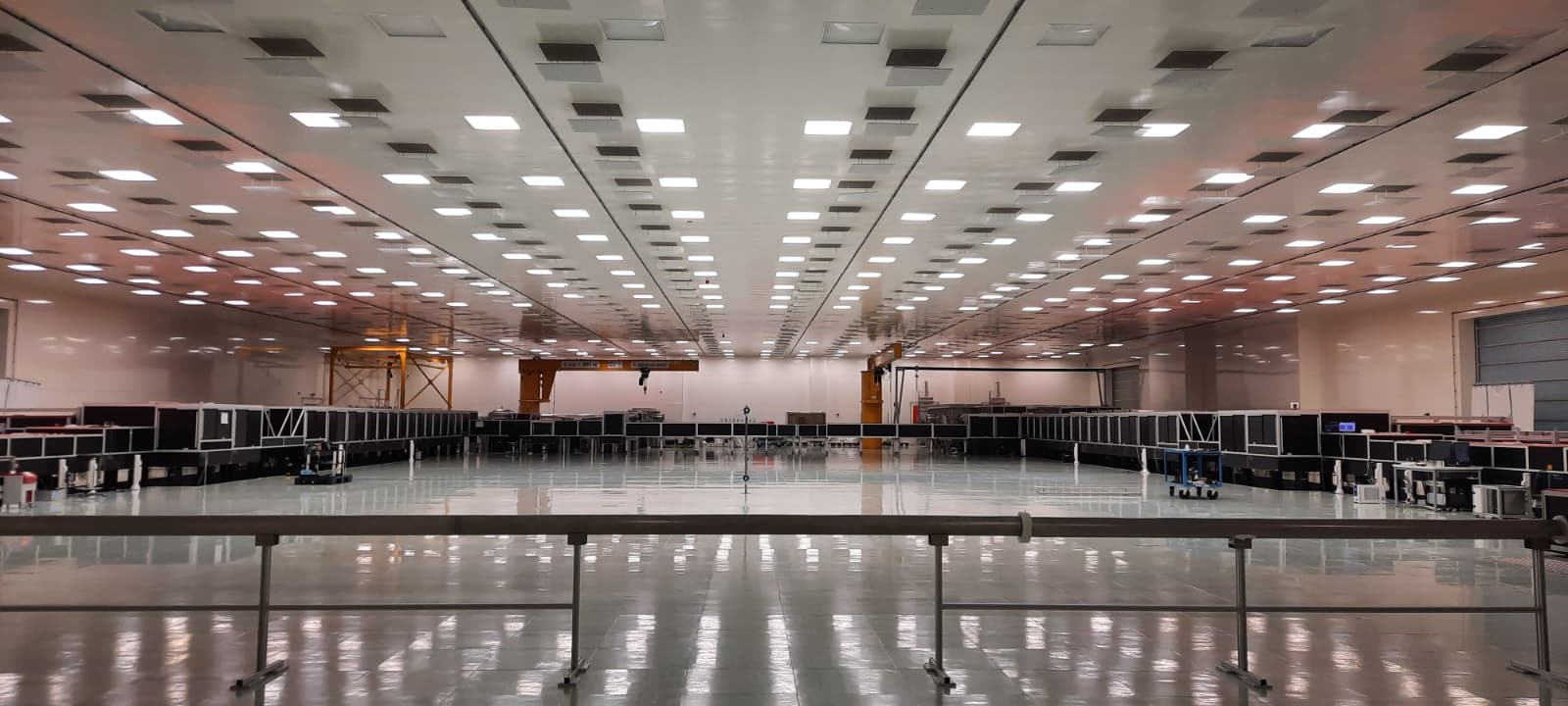}
\caption{Photograph of the High-Power Laser System (HPLS) at ELI-NP. 
Two laser arms for the amplification chains for 0.1PW/1PW/10PW 
output power are arranged on both sides of the photo.}
\label{fig_HPLS}
\end{figure}

\subsection{Vacuum control system for the gas-origin backgrounds}
\begin{figure}[H]
\centering
\includegraphics[trim=0 140pt 0 0, clip, width=\columnwidth]{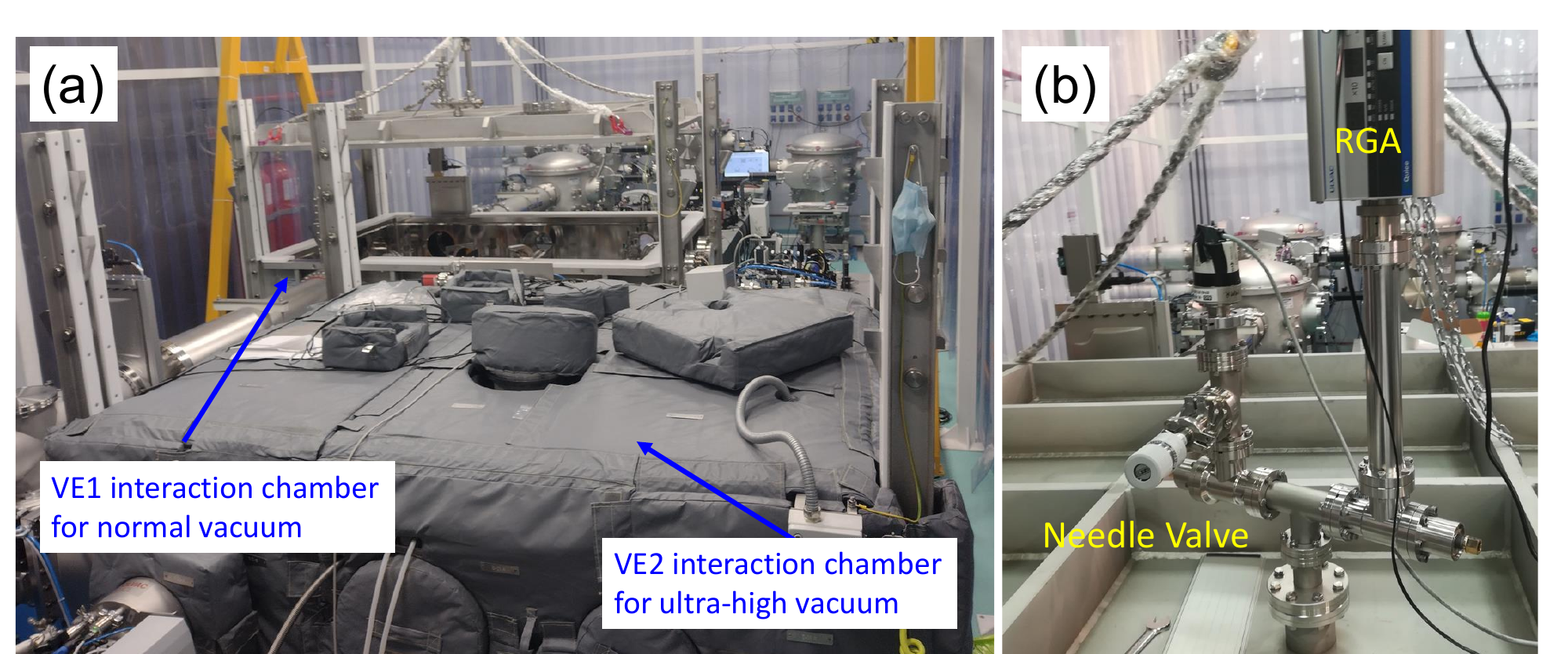}
\caption{\label{fig_vacuum_system_setup} 
(a) The vacuum enclosures for different vacuum-pressure level at the E4 experimental area for 0.1 PW laser output 
at ELI-NP. VE1 and VE2 are dedicated to creating normal vacuum conditions at $5 \times 10^{-7}$ mbar 
and ultra-high-vacuum conditions 
at $5 \times 10^{-9}$ mbar, respectively.
(b) The needle valve and Residual Gas Analyzer (RGA) installed on the lid of the VE1 interaction chamber.}
\end{figure}

The vacuum system is implemented to eliminate the gas-origin background $n_{gas}$ (See Eq. (\ref{eq2})) 
and to study the scaling properties of the pressure dependence. 
Figure \ref{fig_vacuum_system_setup} (a) shows the interaction chambers installed in the E4 experimental area 
for 0.1 PW laser output. 

The vacuum system consists of five turning boxes (TB) and two interaction vacuum enclosures (VE).
VE1 is used as the interaction chamber under normal vacuum conditions in the range 
of $5 \times 10^{-7}$ - $5 \times 10^{-6}$ mbar, which is required for laser propagation 
with minimal nonlinear effects to maintain the spatial coherence over long propagation distances 
from the HPLS to the experimental area. 
The VE2 chamber, which is covered by a baking system in Fig. \ref{fig_vacuum_system_setup}, 
is designed to achieve ultra-high-vacuum condition down to $5 \times 10^{-9}$ mbar. 
All the TBs and beam pipes are constructed according to the ISO-K standard. 
Fluororubber O-rings are used for vacuum sealing. 
In contrast, VE1 and VE2 follow the CF-F standard, employing metallic copper (or silver-plated copper) gaskets 
to minimize the leakage, outgassing and permeation. 
An exception is made for the lids of VE1 and VE2, where fluororubber O-rings are used for easy access to 
the optiomechanical components during the beamtime. 
Alternatively, a double O-ring system with a small channel between the O-ring seals is applied. 
This double O-ring configuration plays a role as a differential pumping stage in order 
to reduce seal permeation by pumping the channel at $5\times10^{-2}$ mbar. 
The vacuum condition is maintained by combining turbomolecular pumps (TMPs) as main vacuum pumps 
and roughing pumps (RPs) for backing pressure. 
TMPs are mounted directly on the vacuum chambers and connected to RPs located in the basement. 

A Residual Gas Analyzer (RGA) and a needle valve were installed on the top of VE1 as shown 
in Fig.\ref{fig_vacuum_system_setup} (b). 
The RGA is used to measure the molecular composition of the residual vacuum in the interaction chamber.
The RGA can provide detailed understanding of the scaling curves related to the gas-origin background ($n_{gas}$).
For instance, FWM generation can be primarily driven by residual water vapor (H$_2$O), 
which is a known polar molecule and a dominant species in residual gases within the targeted pressure range.  

\begin{figure*}[t]
\centering
\includegraphics[scale=0.7]{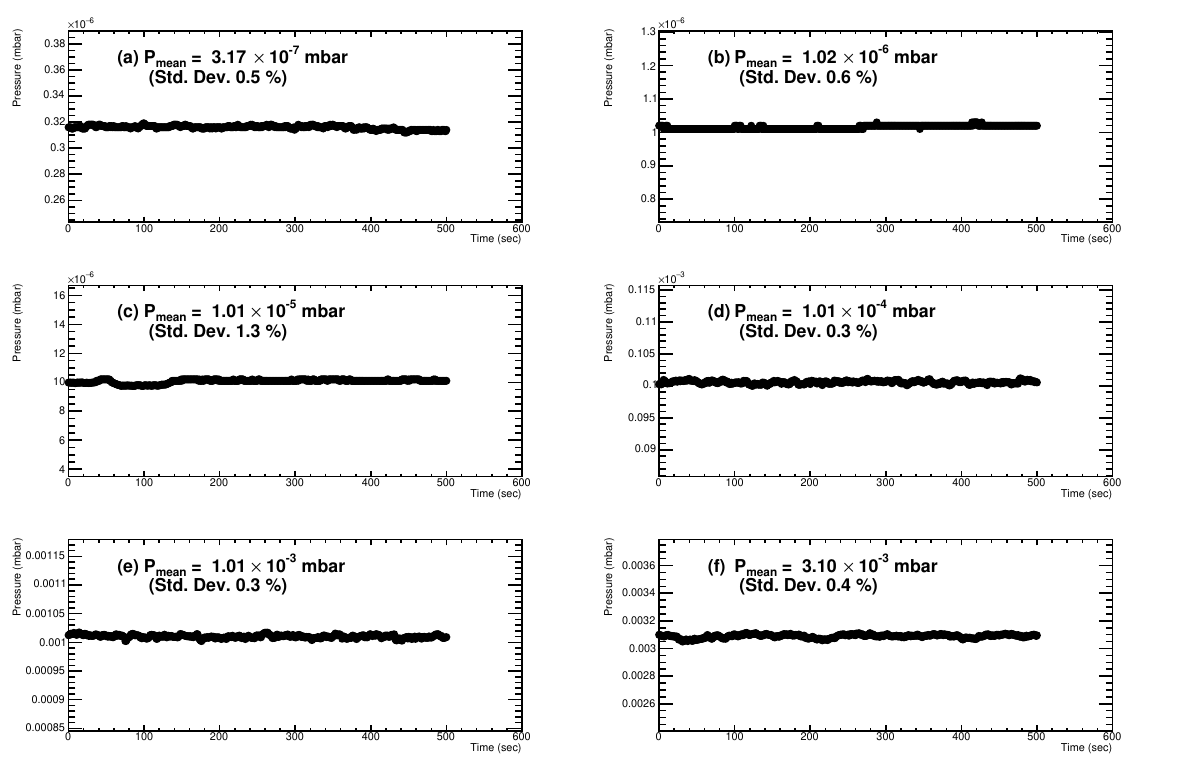}
\caption{Stability of the vacuum pressure as a function of time in the VE1 interaction chamber.
The controlled pressure levels are (a) $3.2\times10^{-7}$, (b) $1.0\times10^{-6}$, (c) $1.0\times10^{-5}$,
(d) $1.0\times10^{-4}$, (e) $1.0\times10^{-3}$, and (f) $3.1\times10^{-3}$ mbar, respectively. 
Each pressure level is controlled by a needle valve.}
\label{fig_vacuum_system_pressure2}
\end{figure*}

The needle valve is used to inject pure nitrogen (or ambient air in the experimental area).
This operation varies the vacuum pressure as well as stabilize it at a desired setpoint 
to study the pressure dependence of the gas-origin background photons. 
The stability of pressure control was measured for 500 seconds (equivalent to 5000 laser-shot events in data acquisition) 
in the range from $3\times10^{-7}$ mbar to $3\times10^{-3}$ mbar, as shown in Fig. \ref{fig_vacuum_system_pressure2}. 
The pressure was stabilized with a standard deviation of 1.3 \% across all the measured pressure levels.

\subsection{Optical system for ALPs search}
\begin{figure*}[t]
\centering
\includegraphics[scale=0.52]{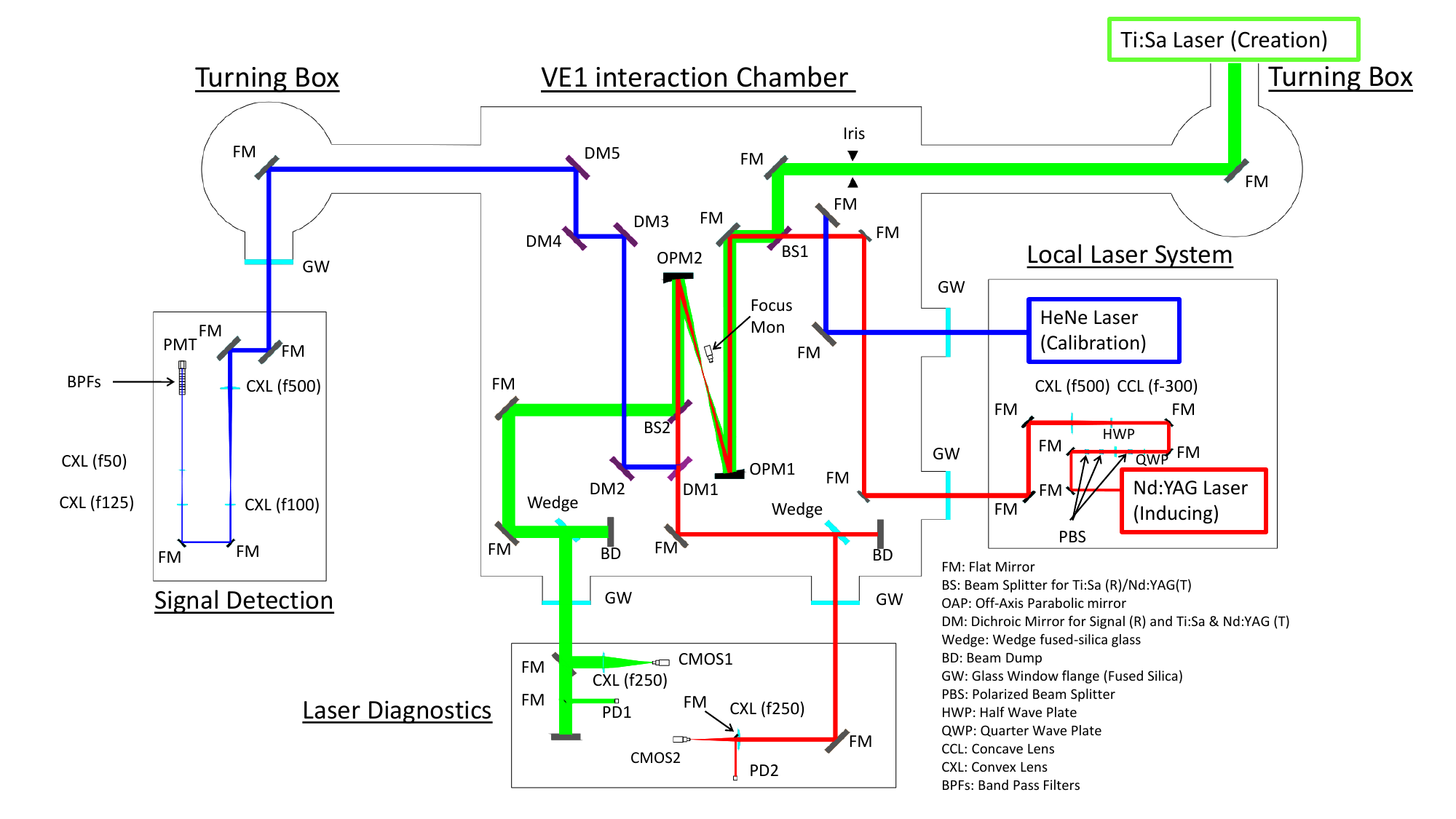}
\caption{Optical setups installed at the E4 experimental area for 0.1 PW laser output at ELI-NP.}
\label{fig_opticalsetup_E4}
\end{figure*}

Figure \ref{fig_opticalsetup_E4} shows the optical setup for the ALPs search with 0.1 PW laser system at ELI-NP.
Ti:Sa creation laser propagates from the HPLS to the VE1 interaction chamber from the right side of Fig. \ref{fig_opticalsetup_E4}.
A motorized iris diaphragm is implemented at the entrance of VE1 to study the area-size dependence 
to identify the optics-origin backgrounds ($n_{opt}$ in Eq.\ref{eq3}) via scaling properties.  

A Nd:YAG laser system was installed on the optical table beside the VE1 interaction chamber. 
The laser system outputs up to 3 J of pulsed energy at the fundamental wavelength of 1064 nm 
with a repetition rate of 10 Hz. 
The energy and polarization state are determined by polarized beam splitters (PBS) and a quarter-wave plate (QWP). 
Polarization control is necessary to determine the polarization-dependent coefficient between pseudo-scalar (e.g., ALPs ) 
and scalar (e.g., dilaton) fields to derive exclusion limites in the theoretical model. 
Nd:YAG laser beam is expanded by a Keplerian telescope composed of a concave lens (CCL) and a convex lens (CVL).
The magnification factor of the telescope is designed such that the diameter of Nd:YAG laser is one-third that of Ti:Sa laser.
This factor is selected to maintain the spatial overlap between the two lasers at the focal spot against 
intrinsic pointing instability of Ti:Sa laser during the data acquisition. 

Ti:Sa creation laser and Nd:YAG inducing laser are combined at BS1 and simultaneously focused by OPM1 
to propagate to the focal point 
for the FWM generation through the ALP production. 
The generated FWM signal, which is expected to be distributed around 641 nm, 
is collimated by the OPM2 and the signal is split from the residual lasers by two beam splitters 
(BS2 and DM1 in Fig.\ref{fig_opticalsetup_E4}). 
BS2 reflects Nd:YAG laser and transmits both Ti:Sa laser and the FWM signal, while DM1 reflects the FWM signal and transmits Ti:Sa laser. 
Two-step separation as such is necessary to suppress the additional production of the optics-orgien FWM 
backgrounds ($n_{opt}$) caused by the two lasers commonly transmitting inside the beam splitter. 
The FWM signal propagates to the detection point through the multiple wavelength selections (DM2 - DM5).
The FWM signal passes through other telescopes to change the collimation size comparable with the clear 
aperture of the single-photon sensitive photomultiplier (PMT) located at the end of the setup 
("Signal Detection" in Fig. \ref{fig_opticalsetup_E4}).
A stack of band pass filters (BPFs) is mounted on the PMT for further elimination of unwanted backgrounds 
at different wavelengths.
The wavelengths around 540 - 710 nm are allowed to be passed through the BPF stack.  
The rejection power for the background photons is OD $ \sim 25$ above 710 nm, mainly against 
the backgrounds in the wavelength range of the residual lasers (800 and 1064 nm), and OD $ \sim 30$ below 540 nm, 
mainly against any possible high-frequency generation from the lasers in that wavelength range, 
according to the manufacturer's data sheets. On the other hand, the acceptance-efficiency of 
FWM signal photons is estimated to be 35 $\%$ including the reflectivities of DMs and the transmittances 
of the BPF stack. 
The acceptance-efficiency was measured by the laser powers at the focal point in VE1 and at Signal Detection 
(PMT in Fig.\ref{fig_opticalsetup_E4}) by using a HeNe calibration laser at a reference wavelength of 632.8 nm.

\begin{figure}[h]
\centering
\includegraphics[width=\columnwidth]{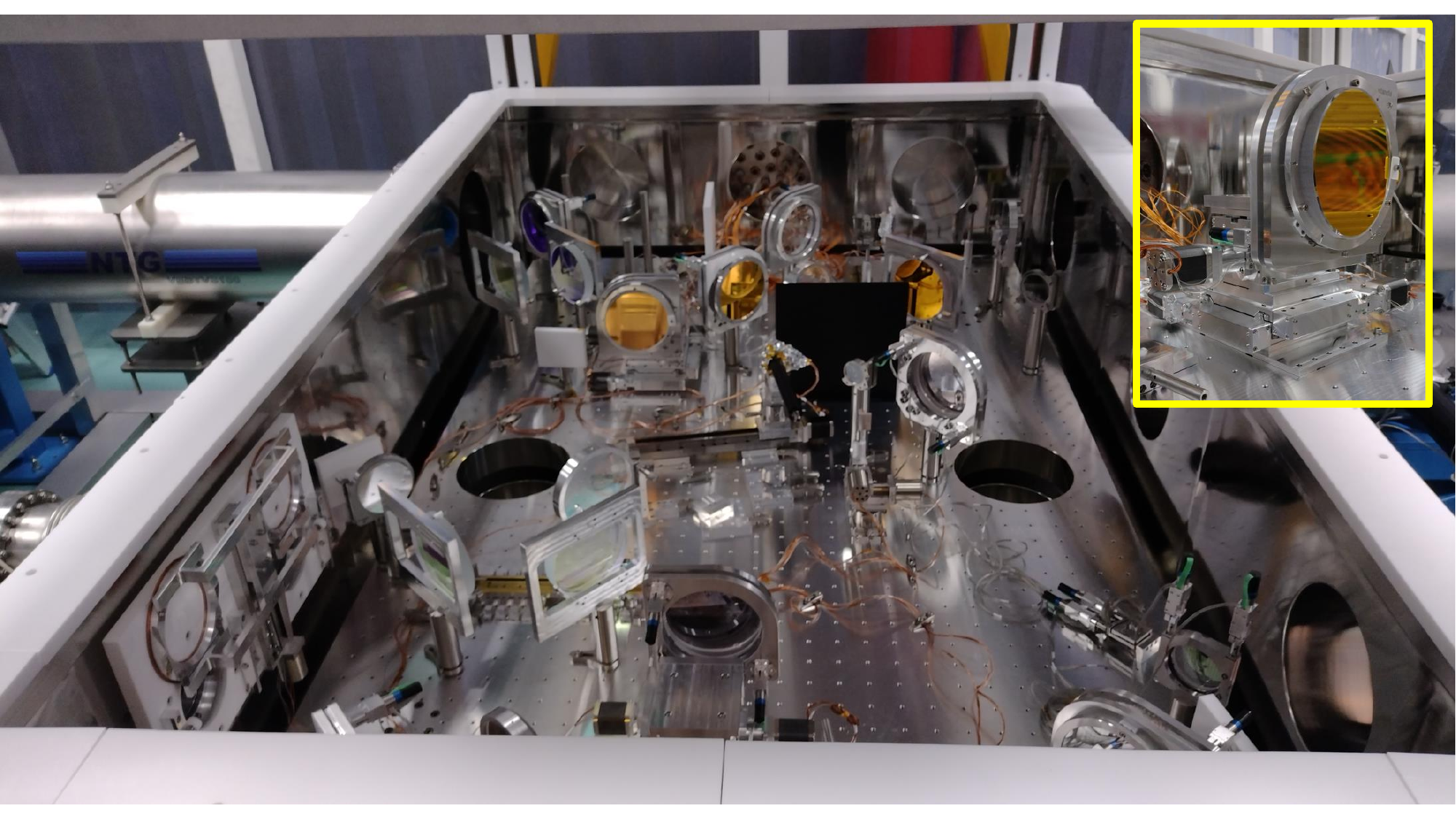}
\caption{Photograph of the experimental setup inside the VE1 interaction chamber. 
The inset photo shows the off-axis parabolic mirror with 5 axes motorized operation 
to handle the focusing and collimating the lasers.}
\label{fig_photo_VE1}
\end{figure}

\begin{figure}[h]
\centering
\includegraphics[width=\columnwidth]{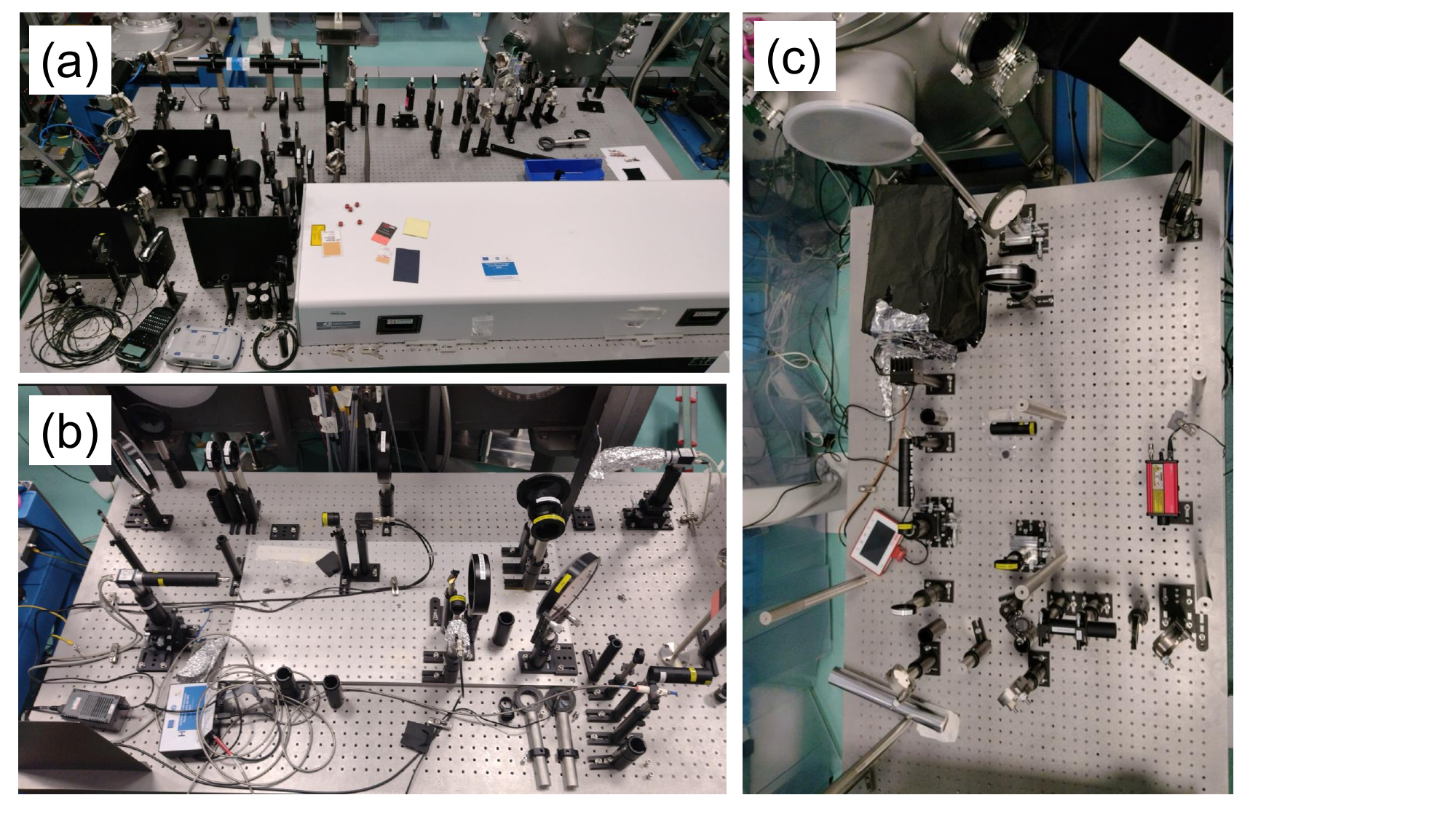}
\caption{Photograph of the experimental setup outside the VE1 interaction chamber. 
 (a) Setup of Nd:YAG inducing laser and HeNe calibration laser. 
 (b) Setup of the laser diagnostics for shot-by-shot monitoring for the spatiotemporal overlap betweent Ti:Sa and Nd:YAG laser pulses. 
 (c) Setup of the detection system for FWM photons observation.}
\label{fig_photo_outsideVE1}
\end{figure}

The two laser beams after BS2 and DM1 are guided to the laser diagnostics. 
At the laser diagnostics, the pointing stability (CMOS1 and CMOS2) and the timing synchronization (PD1 and PD2) 
are monitored on the shot-by-shot basis during the data acquisition 
for a quality control of the laser-shot events in terms of spatiotemporal overlap for the offline data analysis.
Figure \ref{fig_photo_VE1} and \ref{fig_photo_outsideVE1} shows the installation results of optical setups 
for the ALPs search. 

\subsection{Area-size control system for the optics-origin backgrounds and laser diameter measurement}
Fine control of the area size of the laser pulses is crucial to understand scaling properties 
of the optics-origin backgrounds
as a function of laser diameter, distinguishing them from the gas-origin background.
The motorized iris diaphragm (8MID98V-4-H, Standa) was installed into the optical setup, and  
its functionality was tested via laser diameter measurements based on the knife-edge technique. 

\begin{figure*}[t]
\centering
\raisebox{20pt}{\includegraphics[scale=0.22]{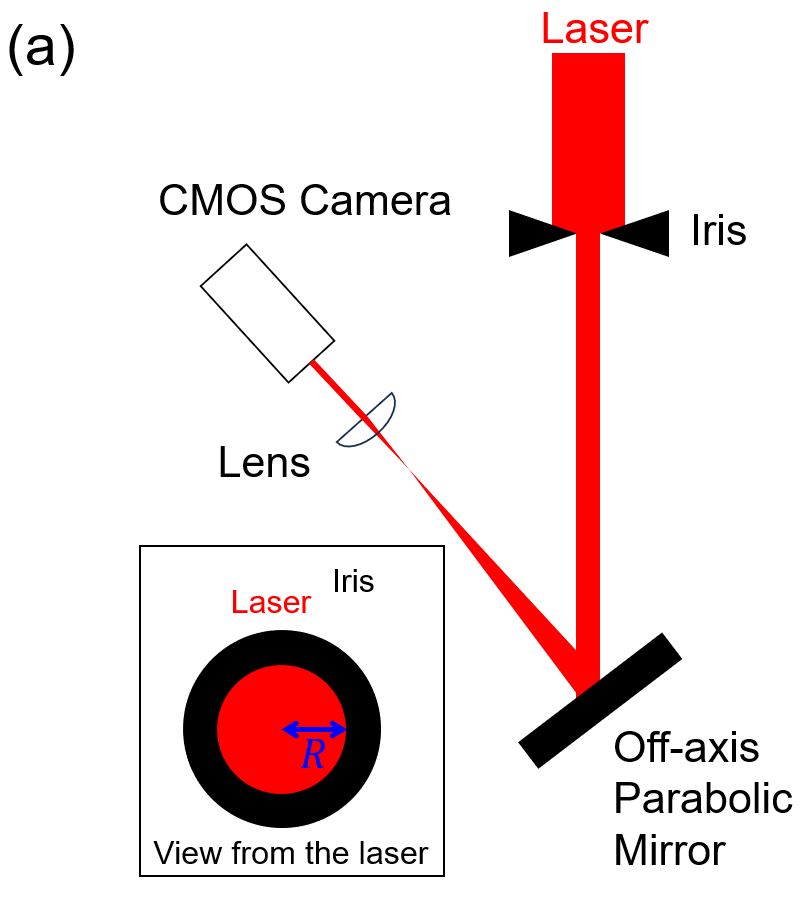}}
\includegraphics[scale=0.24]{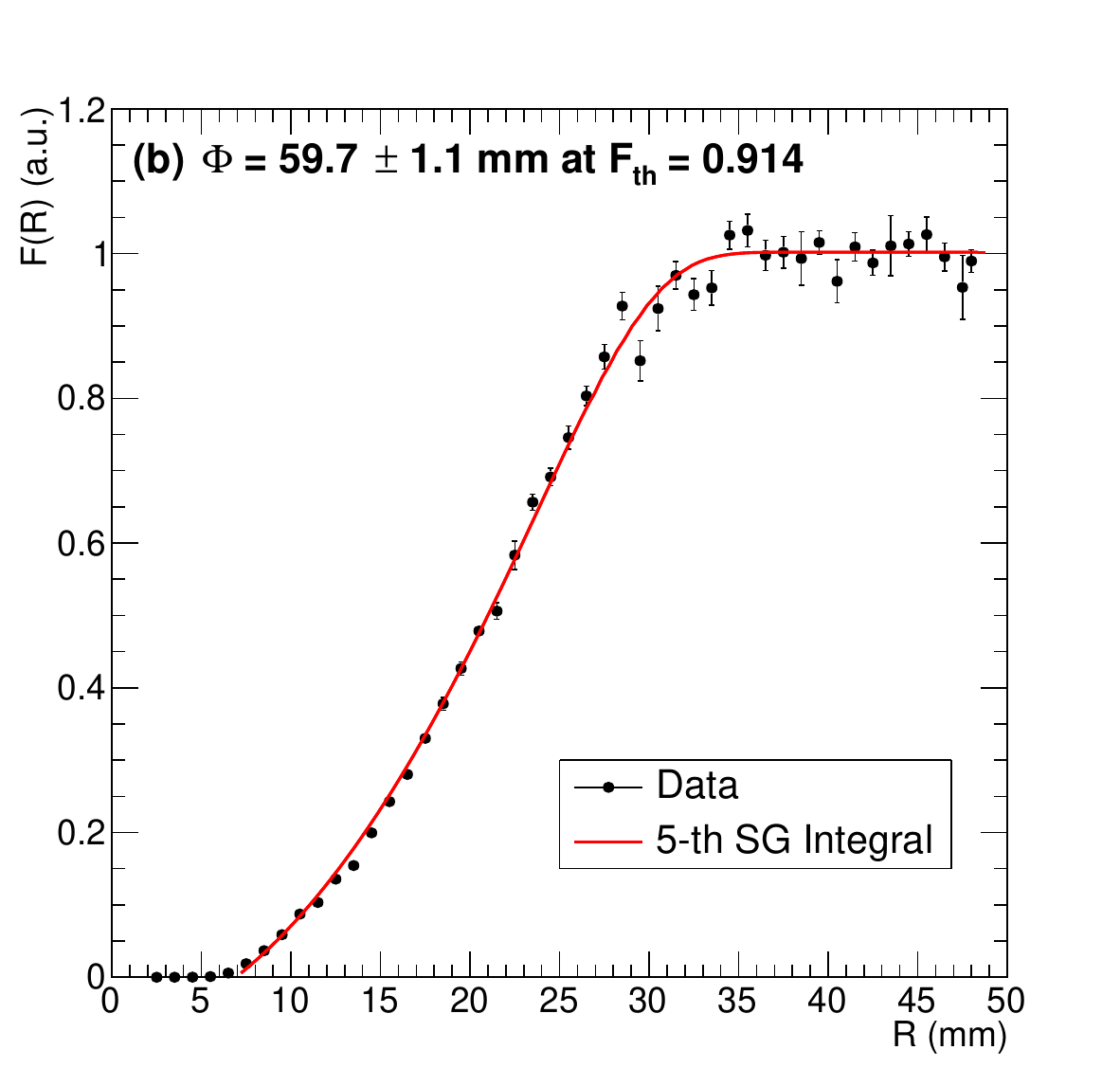}
\includegraphics[scale=0.24]{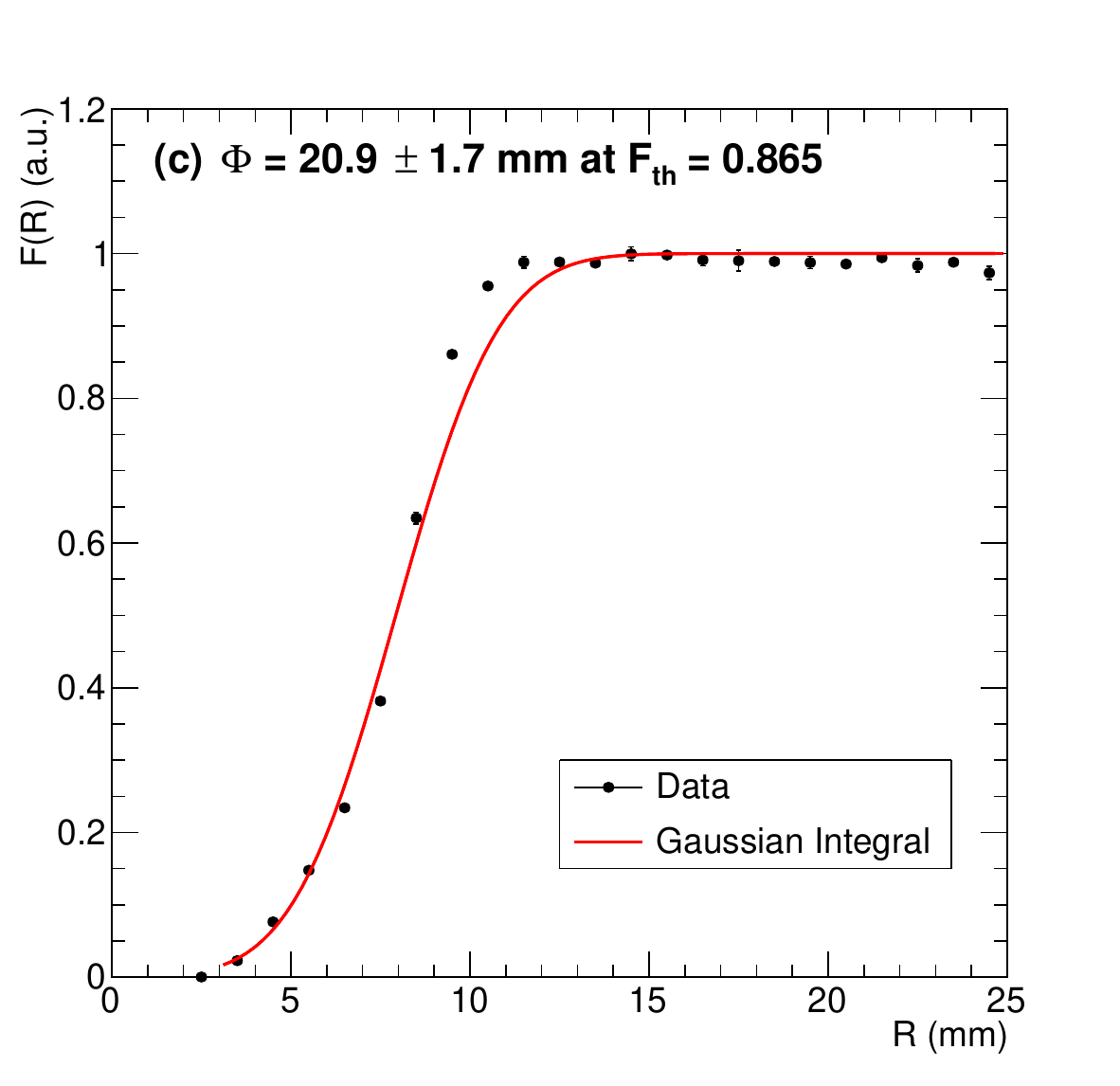}
\caption{(a) Schematic setup for the knife-edge measurement using a motorized iris diaphragm.
(b) Normalized integrated number of photons as a function of the controlled radius $R$ for Ti:Sa laser. 
The diameter of Ti:Sa laser is determined at half maximum at $F(R) = 0.914$.
(c) Normalized integrated number of photons as a function of the controlled radius $R$ for Nd:YAG laser. 
The diameter of Nd:YAG laser is determined at $e^{-2}$ at $F(R)=0.865$.}
\label{Fig_Diameter_IP}
\end{figure*}

Figure \ref{Fig_Diameter_IP} (a) shows the conceptual setup of laser diameter measurement by the knife-edge technique. 
Since Ti:Sa creation laser and Nd:YAG inducing laser are circularly symmetric beams, 
the motorized iris diaphragm was placed before the OPM to execute the knife-edge scan by cutting the beam 
in polar coordinates. 
The CMOS camera, which is used for the "Focus Mon" in Fig.\ref{fig_opticalsetup_E4}, is also utilized 
to measure the integrated number of photons existing within the controlled area size of the laser beam.

The observable of the knife-edge measurement in polar coordinates is defined by 
\begin{eqnarray}
  S\left(R\right) = \frac{\int_{0}^{2\pi} \int_{0}^{R} f\left(r,\theta\right) r dr d\theta } { \int_{0}^{2\pi} \int_{0}^{\infty} f\left(r,\theta\right) rdrd\theta }, 
\end{eqnarray}
where the denominator of the equation indicates the integrated number of photons measured bythe CMOS camera 
when the iris is fully opened. 
In contrast, the numerator indicates the case when the iris is closed to a radius $R$. 
Since the photons are measured in a pixel array on the CMOS camera, the observable is practically expressed by  
\begin{eqnarray}
  F\left( R \right) &=& \frac{\sum_{i=1}^{i=i_{max}}\sum_{j=1}^{j=j_{max}} f\left(x_i,y_j\right)}{F_{norm}},
\end{eqnarray}
where $i$ and $j$ denote the indices of pixels in the horizontal and vertical direction, respectively. 
$f (x_i, y_j)$ denotes the output from each pixel. The coordinates $\left( x_i,y_j \right)$ show 
the relative position with respect to the center of the laser beam. 
The maximal indices $\left(i_{max},j_{max}\right)$ are defined to be $r_{max}=\sqrt{x_{i_{the}}^{2} + y_{i_{th}}^{2}}$, where 
$r_{max}$ is the effective maximal radius of the integration range on the CMOS sensor 
without the fiducial region at the edge of the sensor. 
The normalization factor $F_{norm}$ is determined from the curves of the data points as a function of 
the iris radius $R$ by extrapolating to the limit $R\rightarrow \infty$.

Figure \ref{Fig_Diameter_IP} (b) shows the results of the iris knife-edge measurement for Ti:Sa creation laser. 
The full-aperture diameter of Ti:Sa creation laser is defined at half maximum (HM) assuming the 5th-order Supper-Gaussian beam profiles.
The data points were fitted by an integration form of a 5th-order Super-Gaussian in polar coordinates, that is,  
\begin{eqnarray}
  G\left( R \right) = A\int_{0}^{R}re^{-\left(\frac{r-r_{0}}{\sqrt{2}\sigma}\right)^{2m}},
\label{eq:Dia_fit}
\end{eqnarray} 
where $A, r_0, \sigma$ are the fitting parameters, $m$ is the order of Super-Gaussian. We fixe to $m=5$. 
As a consequence, the diameter of Ti:Sa creation laser is $59.7 \pm 1.1 $ mm 
by solving the inverse function of the fitting such that $R = G^{-1}\left(0.914\right)$.
The value 0.914 represents the fraction of the integrated number of laser photons contained 
in the area at HM for a 5th-order Super-Gaussian beam.   

Figure \ref{Fig_Diameter_IP} (c) shows the results for Nd:YAG inducing laser. 
The full-aperture diameter of Nd:YAG inducing laser is defined at $e^{-2}$ assuming a Gaussian beam, 
following the nominal definition of laser waist size. 
The fitting function is defined at $m=1$ to make Eq.\ref{eq:Dia_fit} express a Gaussian beam, 
where $A, r_0, \sigma$ are the fitting parameters. 
The diameter of Nd:YAG inducing laser is $19.9 \pm 1.6 $ mm by solving the inverse function 
$R=G^{-1}\left(0.865\right)$, which represents the fraction of the integrated number of laser photons
within the laser waist of a Gaussian beam.

Both results regarding the knife-edge observable $F \left( R \right) $ show monotonically increasing trends 
and start to converge to a flat distribution at the full-apperture beam size. 
The results also suggest that the measured diameter of Ti:Sa laser is three times larger 
than that of Nd:YAG laser. It is consistent with the ratio of the spot sizes (although it shows 
an inversely propotional tendency by focusing) of the two lasers,as shown 
by independent measurements of focal-spot size in the next section.

\subsection{Spatial overlap and stabilities}
Ensuring the consistent collision geometry of the two lasers over tens of thousands laser-shot events 
is essential for the interaction of the ALP production with PW-class laser systems 
in order to maintain maximum spaciotemporal overlap factor (Eq. \ref{eq:space-time_overlap}), 
to acquire the data efficiently. In addition, quality control of the focal-spot spatial profile and temporal structure of the high-power laser 
is important in general not only for the ALPs experiment but also for any type of high-power laser experiments. 
Thus, we characterize the quality of the focal-spot image of Ti:Sa creation laser and Nd:YAG inducing laser 
from the viewpoint of pointing stability as well as temporal structure in terms of timing synchronization. 
In this subsection, we discuss the spatial overlap and stabilities, and timing-related issues are discussed in the next subsection.

\begin{figure*}[t]
\centering
\includegraphics[scale=0.252]{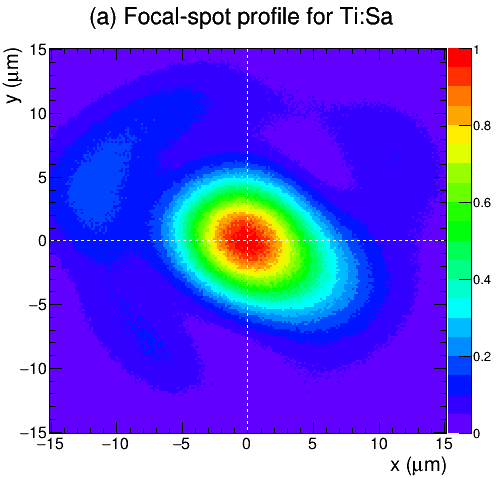}
\includegraphics[scale=0.45]{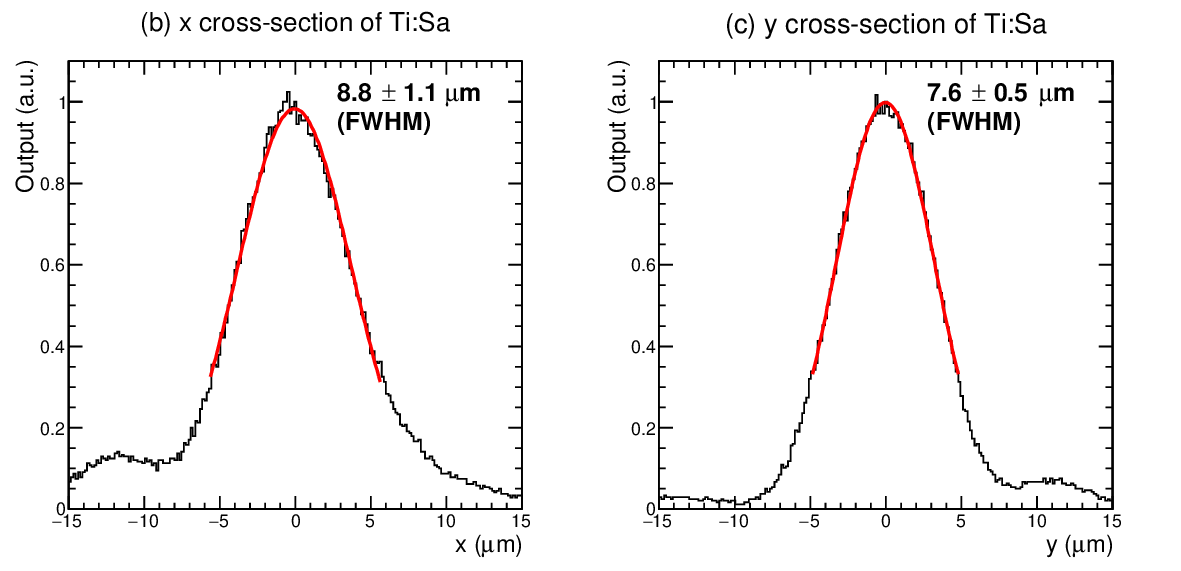}
\includegraphics[scale=0.252]{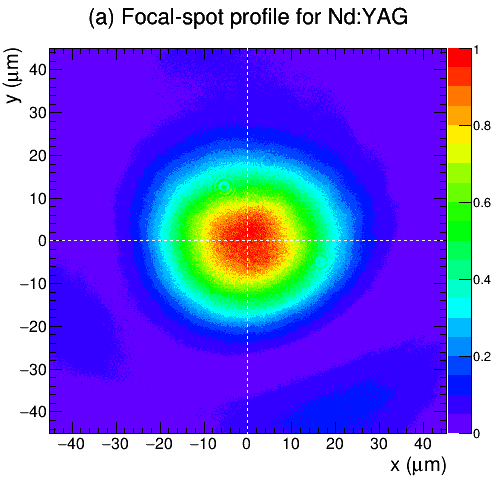}
\includegraphics[scale=0.45]{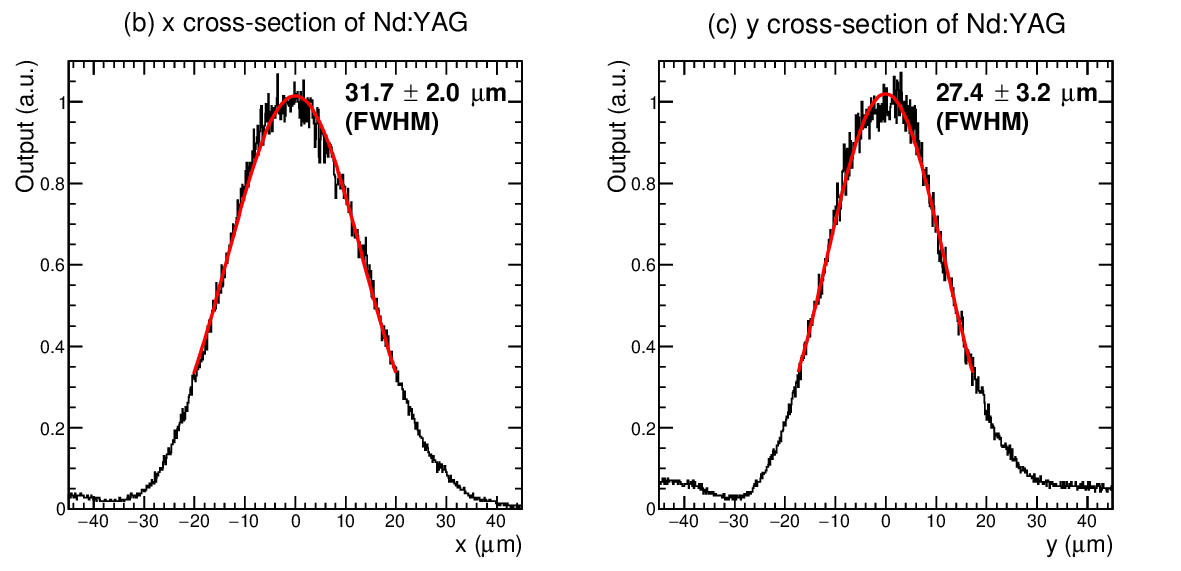}
\caption{(a) Focal-spot image of Ti:Sa creation laser, 
(b) Beam cross-section of Ti:Sa creation laser in $x$, (c) Beam cross-section of Ti:Sa laser in $y$.
(d) Focal-spot image of Nd:YAG inducing laser, 
(e) Beam cross-section of Nd:YAG inducing laser in $x$, (f) Beam cross-section of Nd:YAG inducing laser in $y$.
Dotted while lines on the 2D image addresses the cross-section for $x$ and $y$.}
\label{Fig_FocalSpot_IP}
\end{figure*}

The focal spot images of Ti:Sa creation laser and Nd:YAG inducing laser were directly measured by 
a common CMOS camera placed inside the VE1 interaction chamber (Focus Mon in Fig. \ref{fig_opticalsetup_E4}). 
The camera is remotely controlled to be inserted at the monitoring point and retracted from the optical path 
during the data acquisition for the ALPs search.
Figure \ref{Fig_FocalSpot_IP} (a) and (d) show the 10-shot averaged focal-spot images for Ti:Sa creation laser and Nd:YAG inducing laser, respectively.
The intersection of the two white dotted lines on the 2D images addresses the point of peak intensity of the laser. 
The beam cross-sections along with those lines are also shown as 1D histograms in the other figures.  
The spot size of the beam is evaluated by fitting with a Gaussian function 
in $x$ (horizontal) direction and $y$ (vertical) direction, separately. 
The full width of Ti:Sa laser at half maximum is $8.8 \, \pm 1.1 \, \mu$m for $x$ (Fig.\ref{Fig_FocalSpot_IP} (b)) 
and $7.6 \, \pm 0.5 \, \mu$m for $y$ (Fig.\ref{Fig_FocalSpot_IP} (c)).
In term of Nd:YAG laser, the full widths are  $31.7 \, \pm 2.0 \, \mu$m for $x$ (Fig.\ref{Fig_FocalSpot_IP} (e)) 
and  $27.4 \, \pm 3.2 \, \mu$m for $y$ (Fig.\ref{Fig_FocalSpot_IP} (f)). 
The errors represent the standard deviation of the spot size over 10 laser shots. 
The measured spot size of Nd:YAG laser is about three times larger than that of Ti:Sa laser, as designed.

\begin{figure}[H]
\centering
\includegraphics[scale=0.4]{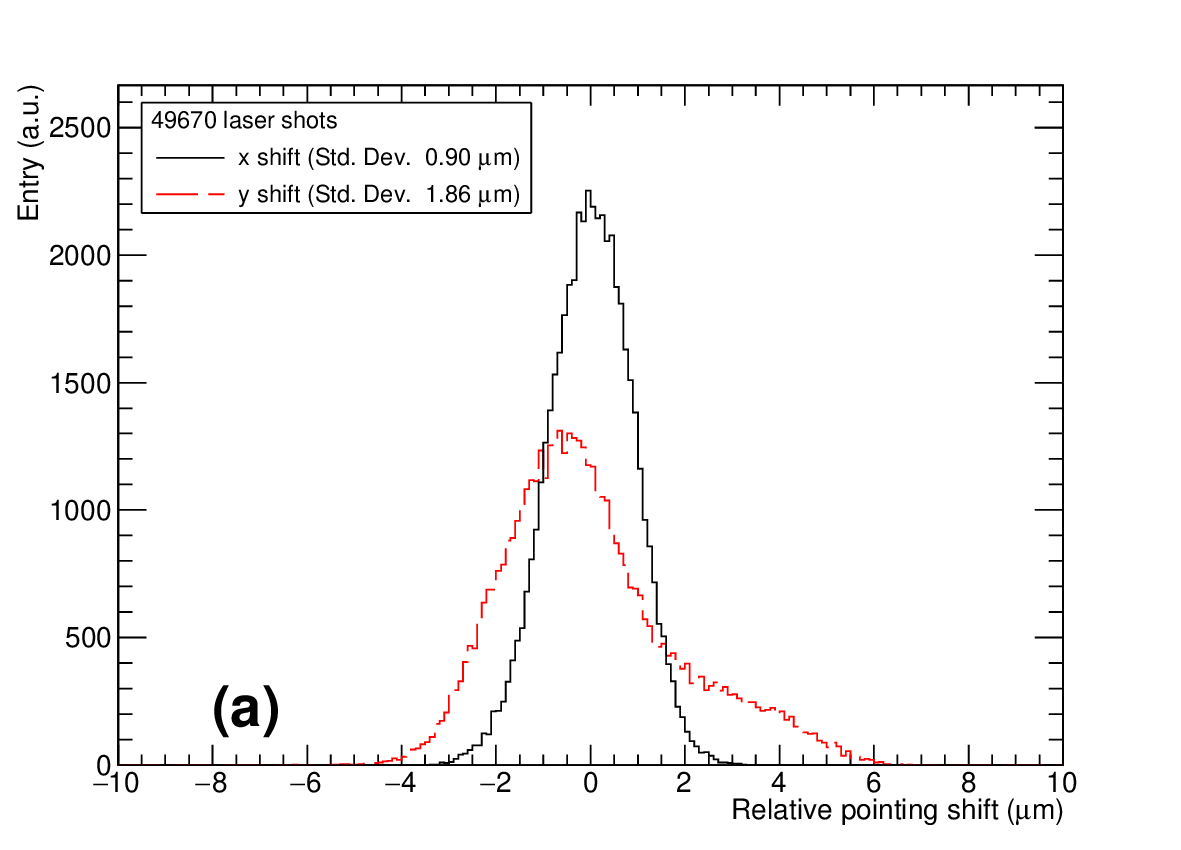}
\includegraphics[scale=0.4]{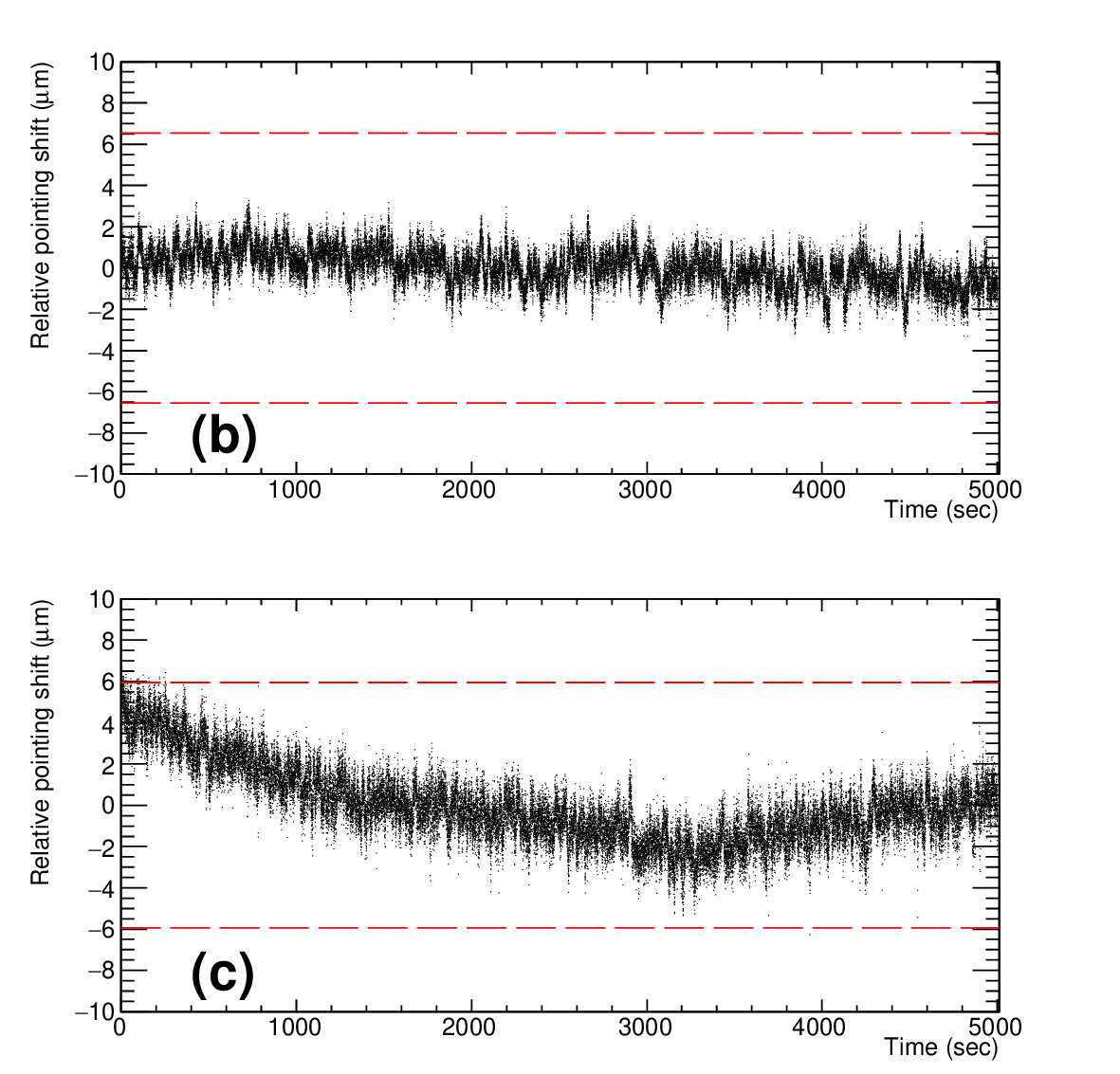}
\caption{\label{Fig_Pointing}
(a) Histogram of the pointing shift of Ti:Sa creation laser-beam 
with respect to the average center of the focal-spot profile. 
The average center is calculated by using 49670 laser-shot data samples. 
Black solid and red dotted curves show the relative pointing shifts
for the $x$ direction (horizontal) and the $y$ (vertical) direction, respectively. 
(b) Time dependence of the relative pointing shift in the $x$ (horizontal) direction. 
(c) Time dependence of the relative pointing shift in the $y$ (vertical) direction. 
Dotted lines in (b) and (c) show a reference of $\pm \, 0.5 \, \sigma$ deviation 
from the center of Nd:YAG inducing laser. 
}
\end{figure}

Pointing stability is evaluated by continuous measurements of the 2D focal-spot profile over 
about 50k laser-shot events. 
Relative pointing shiftw of the lasers are defined as 
\begin{eqnarray}
\Delta x_{i} & = & x_{i}-\mu_{x},\:
\Delta y_{i}   =  {y}_{i}-\mu_{y},\\
\mu_{x} & = & \frac{1}{n_{shot}} \sum\limits_{i=1}^{n_{shot}}x_i,\: 
\mu_{y}   =  \frac{1}{n_{shot}} \sum\limits_{i=1}^{n_{shot}}y_i
\end{eqnarray}
where $\left( x_{i}, y_{i} \right)$ denote the coordinates of the intensity centroid 
of the 2D beam profile at focal spot. The subscript $i$ denotes the laser-shot event number 
and $n_{shot}$ denotes the total number of laser-shot events.  
$\mu_{x}$ and $\mu_{x}$ indicate the average values over 49670 laser-shot events 
for $x_{i}$ and $y_{i}$, respectively. 
Figure \ref{Fig_Pointing} (a) shows the histogram of the pointing fluctuations 
for $x$ (black solid) and $y$ (red dotted). 
The histogram for $x$ shows a symmetric distribution around zero, whereas that for $y$ 
shows an asymmetric distribution. The asymmetry suggests that the $y$ shift includes 
a systematic drift of the laser beam over time in addition to random shot-by-shot fluctuations.
Figure \ref{Fig_Pointing} (b) and (c) show the tendency of the pointing shift over time 
to clarify the systematic drift.
The $x$ shift can be explained by shot-by-shot fluctuations with $\sim 1 \, \mu$m standard deviation 
and $\sim 2 \, \mu$m as peak-to-valley deviation. 
The dotted lines in Fig. \ref{Fig_Pointing} (b) shows a reference 
which shows the deviation of $\pm \, 0.5 \, \sigma$ around the intensity-centroid of Nd:YAG laser. 
Thus, the shot-by-shot $x$ fluctuation of a few $\mu$m has little influence on the spatial overlap of the two lasers. 
Whilst, as shown in Fig\ref{Fig_Pointing} (c), the focal-spot position slowly drifts in the $y$ direction 
as the time goes by. 
The drift effect results in $\sim 1.9 \, \mu$m standard deviation, 
which is nearly double that of the $x$ shift. 
Nevertheless, we can conclude that the peak of Ti:Sa creation laser remains within $\pm \, 0.5 \, \sigma$ 
deviation range of Nd:YAG inducing laser, once the spatial overlap of the two lasers is adjusted 
between data aquisitions. 
We emphasize that these results state the worst case of the stability because 
typical necessary statistics are between 4000 and 8000 events (800 seconds at maximum for 10 Hz data acquisition)
as a single data set, and the spatial overlap can be adjusted every several runs, 
according to the online shot-by-shot laser monitoring in the laser diagnostics. 

\subsection{Timing synchronization and stabilities}
The quality control of the temporal shape of the laser pulses are necessary, in particular, 
ensuring the femtosecond pulse duration of 0.1 PW Ti:Sa laser is a must to characterize 
the ultrashort high-power laser system. 
To measure femtosecond pulse durations, we employ Spectral Phase Interferometry 
for Direct Electric Field Reconstruction (SPIDER), which was originally reported in the references \cite{SPIDER1,SPIDER2}.
SPIDER is a powerful method for reconstructing the temporal pulse shape for a few-cycle laser field, 
typically applicable for 5 - 50 fs pulse duration on a single-shot basis.
The technique utilizes the principle of spectral shearing interferometry in the frequency domain 
to determine the relative spectral phase
$\phi \left(\omega\right)-\phi\left(\omega+\Omega\right)$ .
The spectral interferogram is generated by introducing a linear temporal delay $\tau$ 
between two replicas of the original laser pulse and a controlled frequency shift $\Omega$ (i.e., spectral shear) 
produced by a parametric process interacting between the two replicas and 
the third linearly chirped pulse with different instantaneous frequencies.
In a typical configuration of SPIDER, which is well described in the reference \cite{SPIDER3}, 
for example, an original laser pulse at the central wavelength of 800 nm is split into 
two replicas with an optical delay of $\tau$ between them. 
These replicas are up-converted to different frequencies 
inside a nonlinear crystal through the interactions with different temporal portions of the chirped pulse.
In such way, the spectral shear $\Omega$ is produced for the spectral interferogram. 
Consequently, the spectral interferogram is given by 
\begin{eqnarray}
  & & S\left(\omega\right) = I\left(\omega+\Omega\right) + I\left(\omega\right) \nonumber  \\ 
  & & \,\, + \, 2\sqrt{I\left(\omega\right)I\left(\omega+\Omega\right)}
  \cos \left[\phi\left(\omega+\Omega\right)-\phi\left(\omega\right)+\tau\omega\right]
\end{eqnarray}
Let us show the complex form of the spectral interferogram below to clarify the extraction of the relative phase.  
\begin{eqnarray}
  S\left(\omega\right) = S_{0}\left(\omega\right) + S_{-}\left(\omega\right)e^{-i\tau\omega} 
  + S_{+}\left(\omega\right)e^{i\tau\omega},
\end{eqnarray}
where $S_{0} = I\left(\omega+\Omega\right) + I\left(\omega\right)$, and
$S_{\pm} = \sqrt{I\left(\omega\right)I\left(\omega+\Omega\right)} \exp{\left[\pm i\left(\phi\left(\omega+\Omega\right)-\phi\left(\omega\right)\right)\right]}$.
By applying Fourier transform to $S\left(\omega\right)$, the three terms are separated in the time domain at $t=0$ and $t=\pm\tau$ due to the time-shifting property. 
Thus, the relative spectral phase is obtained by selecting the time range around $t=\tau$ with a filter function $H\left( t \right)$ as
\begin{eqnarray}
& & \phi\left(\omega+\Omega\right)-\phi\left(\omega\right)+\tau\omega \nonumber \\
& & \qquad \qquad = \arg \left[S_{+}\left(\omega\right)\right] \nonumber \\
& & \qquad \qquad = \arg \left( \mathcal{F}\left[ \mathcal{F}^{-1}  \left[ S \left( \omega \right) \right] H \left( t \right) \right] \right),
\end{eqnarray}
where the symbols $\mathcal{F}$ and $\mathcal{F}^{-1}$ denote Fourier transform and inverse Fourier transform, respectively. 
The product $\tau\omega$ is determined in the calibration process without spectral shear. 
In the end, frequency-dependent group delay is obtained as an approximation below. 
\begin{eqnarray}
  \frac{d\phi\left(\omega\right)}{d\omega} \approx \frac{\phi\left(\omega+\Omega\right)-\phi\left(\omega\right)}{\Omega}
\end{eqnarray}
The temporal profile of the laser electric field (or intensity) is reconstructed from the measured 
$\frac{d\phi\left(\omega\right)}{d\omega}$ in combination with an independent measurement of the spectral intensity $ I\left(\omega\right)$. 

\begin{figure}[!h]
\centering
\includegraphics[scale=0.32]{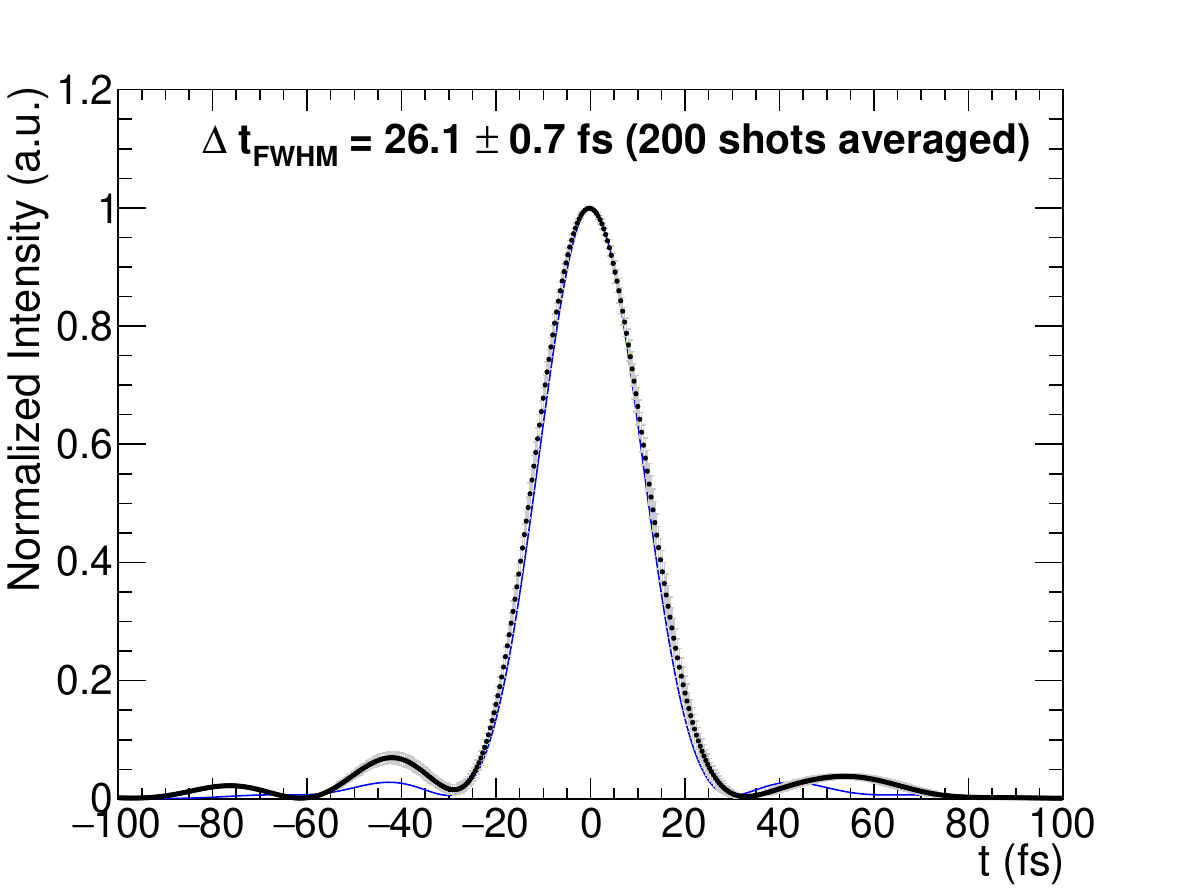}
\caption{\label{Fig_Temporal_TiSa} Temporal intensity profile of Ti:Sa creation laser reconstructed by SPIDER.
The black curve shows the profile averaged over 200 laser-shot events. Gray-shaded band shows the standard deviation. 
The blue solid curve is the Fourier-limited temporal profile calculated by the measured spectrum via Fourier transform.}
\end{figure}

The black curve in Figure \ref{Fig_Temporal_TiSa} shows the reconstructed temporal intensity profile of Ti:Sa creation laser,
averaged over 200 laser-shot events. The gray-shaded band shows the standard deviation of the pulse shape.
The blue solid curve depicts the Fourier-limited temporal profile calculated by Fourier transform of the measured frequency spectrum.
The main peak of the reconstructed temporal shape is in good agreement with the Fourier-limited temporal profile.  
The pulse duration is evaluated to be $26.1 \pm 0.7 $ fs as the full width at half maximum (FWHM). 

\begin{figure}[H]
\centering
\includegraphics[scale=0.32]{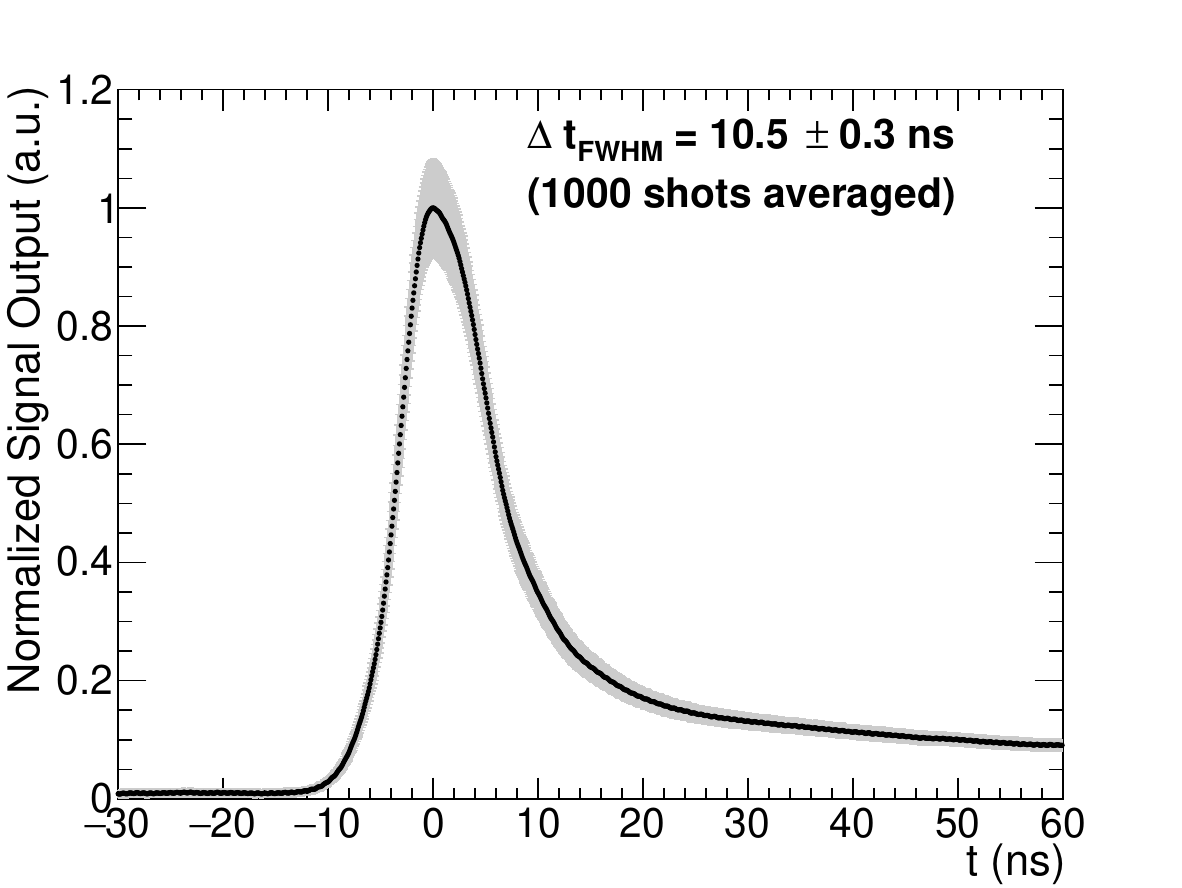}
\caption{\label{Fig_Temporal_NdYAG} Raw temporal distribution of Nd:YAG inducing laser. The measurement was done 
by using fast InGaAs photodiode. The temporal distribution is plotted after averaging over 1000 laser-shot events. 
Gray-shaded band shows the standard deviation of the distribution.
Pulse duration of Nd:YAG inducing laser is $10.5 \pm 0.3$ ns at FWHM.}
\end{figure}

In contrast to Ti:Sa laser, measuring the pulse duration of Nd:YAG laser is more straightforward. 
The temporal profile of Nd:YAG laser is measured using a fast InGaAs photodiode (UPD-35-UVIR-P, Alphalas),
which features a 35 ps rise and fall time. Figure \ref{Fig_Temporal_NdYAG} shows 
Nd:YAG temporal profile averaged over 1000 laser-shot events, 
where the gray-shaded area represents the standard deviation. 
The pulse duration was evaluated at the full-width at half-maximum (FWHM) of the waveform. 
The resulting pulse duration is $10.5 \pm 0.3$ ns at FWHM. 

\begin{figure}[H]
\centering
\includegraphics[width=\columnwidth]{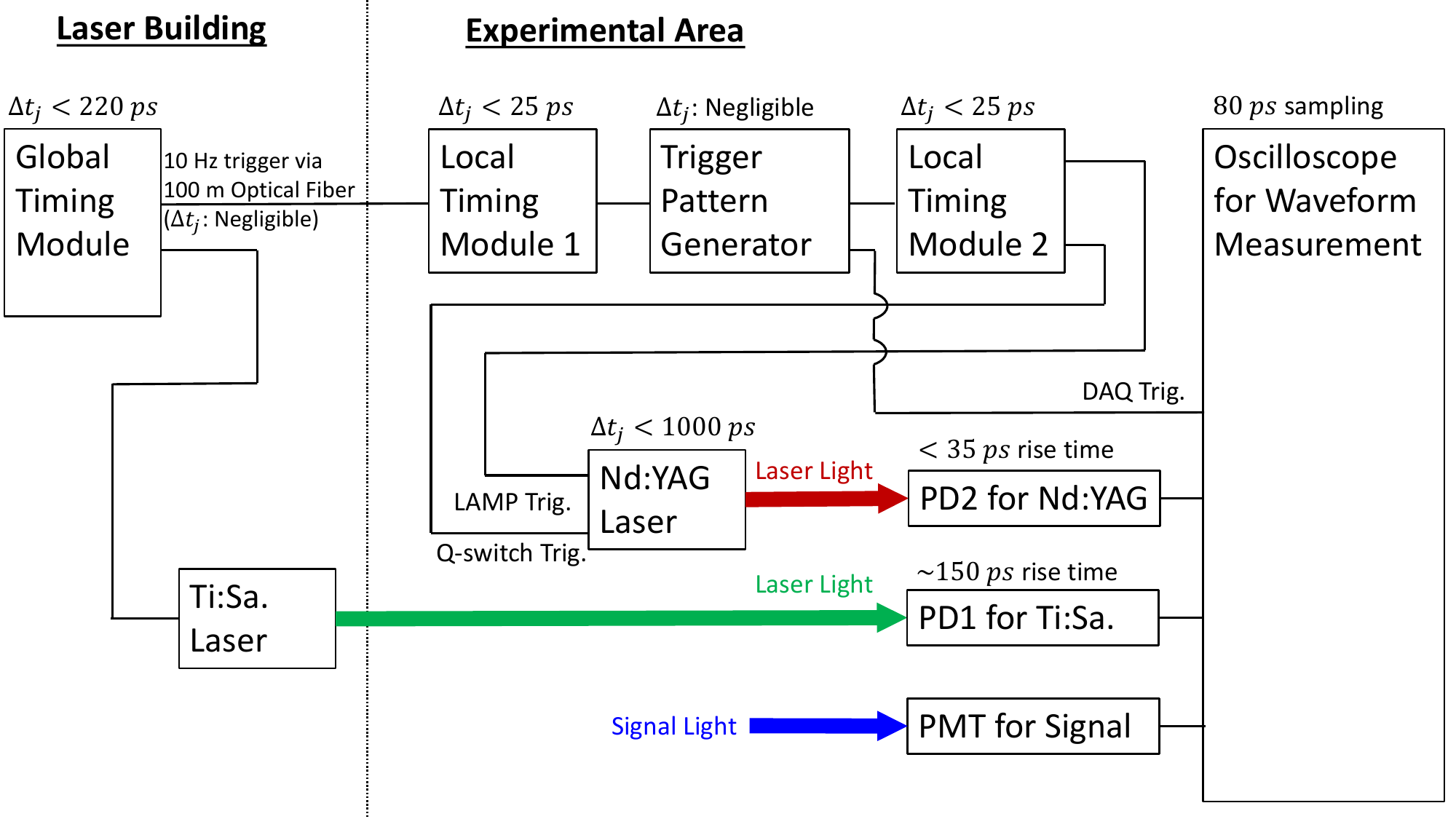}
\caption{Timing control at he E4 experimental area for 0.1 PW laser output at ELI-NP. 
External triggers are delivered to operate Nd:YAG laser for timing synchronization 
and to take the waveform date for four-wave mixing signal via multiple daisy-chain connections of local timing modules. 
A low-jitter timing control system was implemented in the HPLS and the experimental area 
for timing synchronization with fast detection device 
as well as Nd:YAG inducing laser. 
The timing jitter is suppressed to be much less than intrinsic timing jitter of Nd:YAG laser itself. }
\label{fig_trigger_system_setup}
\end{figure}

Timing synchronization between Ti:Sa creation laser and Nd:YAG inducing laser 
can be handled electronically due to the 10 ns pulse duration of Nd:YAG laser, 
provided that Q-switch timing of Nd:YAG laser is determined 
by a low-jitter external trigger. 
Figure \ref{fig_trigger_system_setup} shows a block diagram of the timing control system installed 
at the 0.1 PW experimental area. 
The "Global Timing Module" generates a fundamental timing for 0.1 PW Ti:Sa laser operation. 
A trigger signal output from the module is delivered to the experimental area via an optical fiber 
over a distance of approximately 100 m. 
These optical fibers are employed for the transmission of logic signals from the HPLS to the experimental area 
to prevent additional timing jitter caused by pulse-shape distortion over long distances. 
At the entrance of the experimental area, the optical signal is converted back into a TTL logic signal 
and delivered to Nd:YAG laser system and the oscilloscopes for the waveform data acquisition 
via daisy-chain connections of multiple local timing modules (DG645, SRS).
All the intermediate timing modules utilize a rubidium timebase 
to suppress the timing jitter to less than 25 ps.
Consequently, the total timing jitter introduced by the entire timing control path is significantly 
lower than the intrinsic jitter of Nd:YAG laser itself.

\begin{figure}[H]
\centering
\includegraphics[scale=0.32]{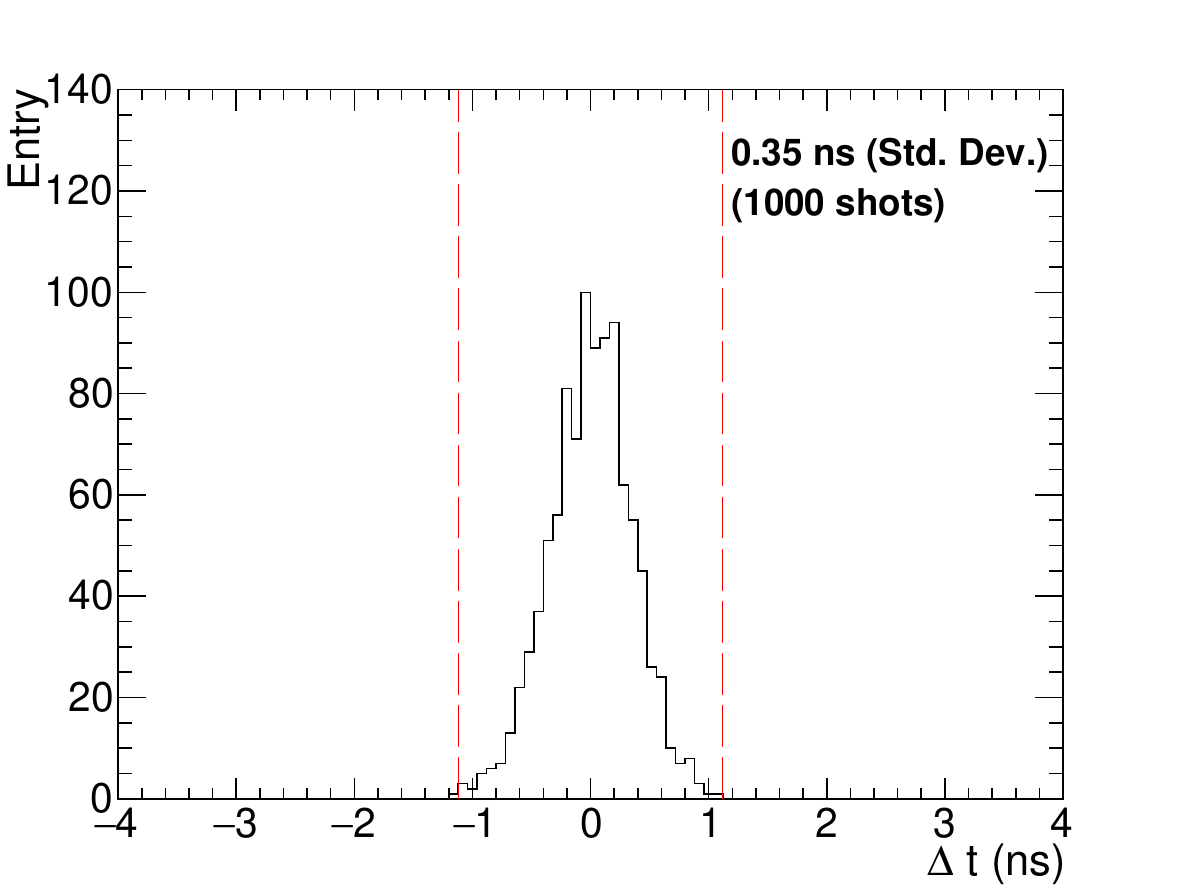}
\caption{\label{Fig_Timing} Timing jitter between Ti:Sa creation laser with respect to Nd:YAG inducing laser
for 800 laser shots.}
\end{figure}

Raw signals from both laser pulses are measured via fast photodiodes on a shot-by-shot basis 
to monitor the timing as well as to diagnose the output status of the laser pulses. 
A Si photodiode with a 150 ps response (DET25A/M, Thorlabs) is used for Ti:Sa laser, 
while a 35 ps response InGaAs photodiode (UPD-35-UVIR-P, Alphalas) is used for Nd:YAG laser. 
The photodiode outputs are recorded as waveforms using a 1 GHz Oscilloscope with an 80 ps sampling interval 
(MSO45, Tektronix). 
The timing jitter between the two lasers is evaluated by identifying the rising edge of the waveforms 
at half maximum for both pulses. Figure \ref{Fig_TimingSync} shows timing jitter of Nd:YAG inducing laser relative to Ti:Sa creation laser, 
evaluated by using 800 data samples. 
The measured timing jitter is $\pm$ 350 ps (standard deviation), which is well suppressed 
compared to the $\pm \,0.5 \sigma$ equivalent values (dotted lines) of the Nd:YAG pulse duration.

\subsection{Triggered event selection for background source separation}
\begin{figure}[H]
\centering
\includegraphics[width=\columnwidth]{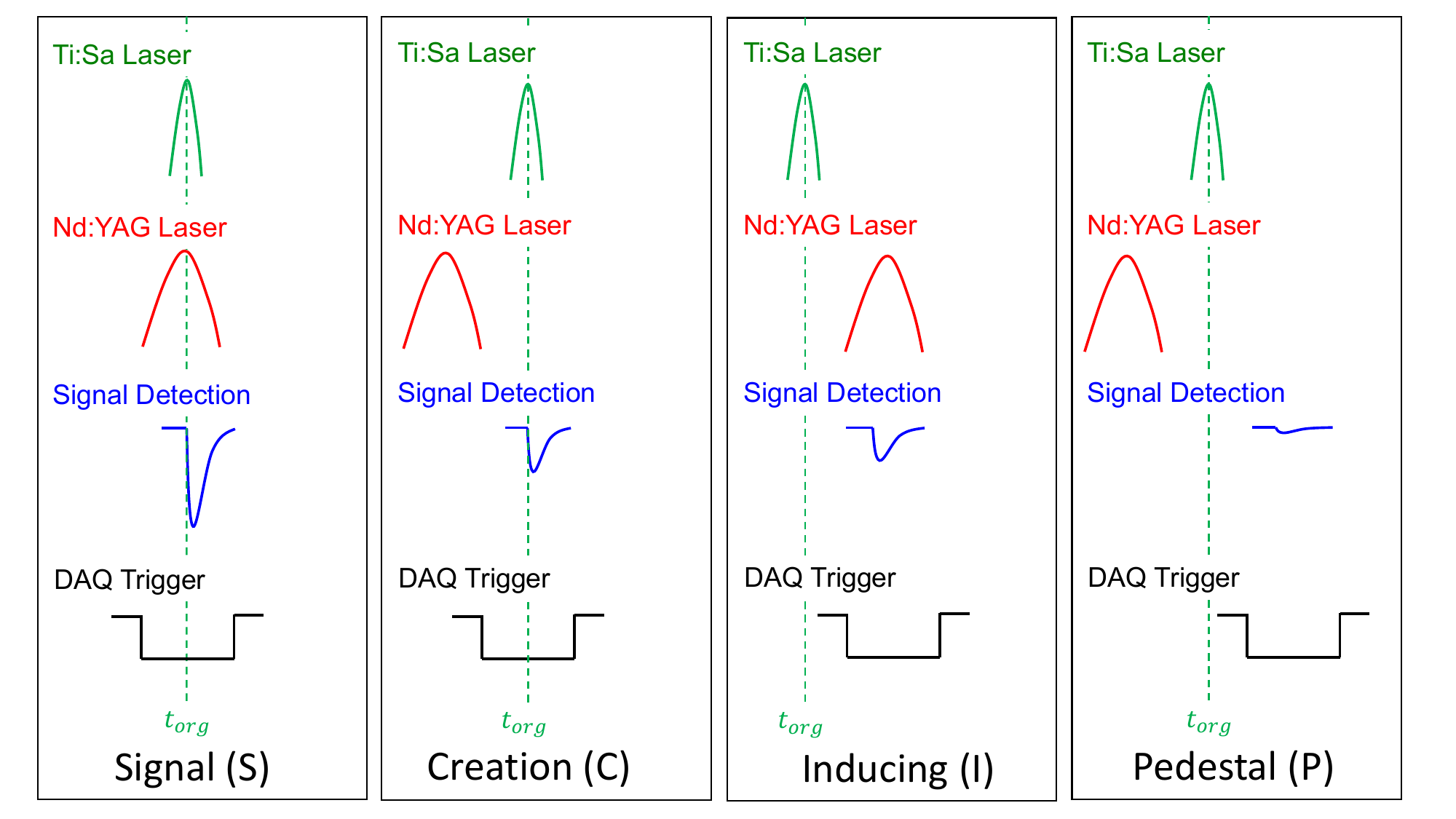}
\caption{\label{Fig_trigger_pattern} The schematic illustration shows synchronization (or desynchronization) 
of the timings for Ti:Sa creation laser, Nd:YAG inducing laser, and the start timing of 
the waveform data acquisitions in order to generate the trigger patterns "Signal (S)", "Creation (C)", "Inducing (I)" and "Pedestal (P)".}
\end{figure}

A Trigger Pattern Generator (TPG) is implemented in the timing control system 
(Fig. \ref{fig_trigger_system_setup}). 
The purpose of the TPG is to create four distinct trigger states across every four laser shots, 
defined as "Signal (S)", "Creation (C)", "Inducing (I)", and "Pedestal (P)". 
Figure \ref{Fig_trigger_pattern} schematically illustrates these patterns with the expected timing shifts 
of the laser pulses, the FWM signal, and the time window of the data acquisition.
In the "S" pattern, FWM signals can be detected with the timing synchronization between both lasers 
whithin the time window of the data acquisition. 
Conversely, the "C" and "I" patterns detect background contributions from only one of the lasers. 
The "P  " pattern provides the pedestal noise detection independent of the laser timing.

\begin{figure}[H]
\centering
\includegraphics[width=\columnwidth]{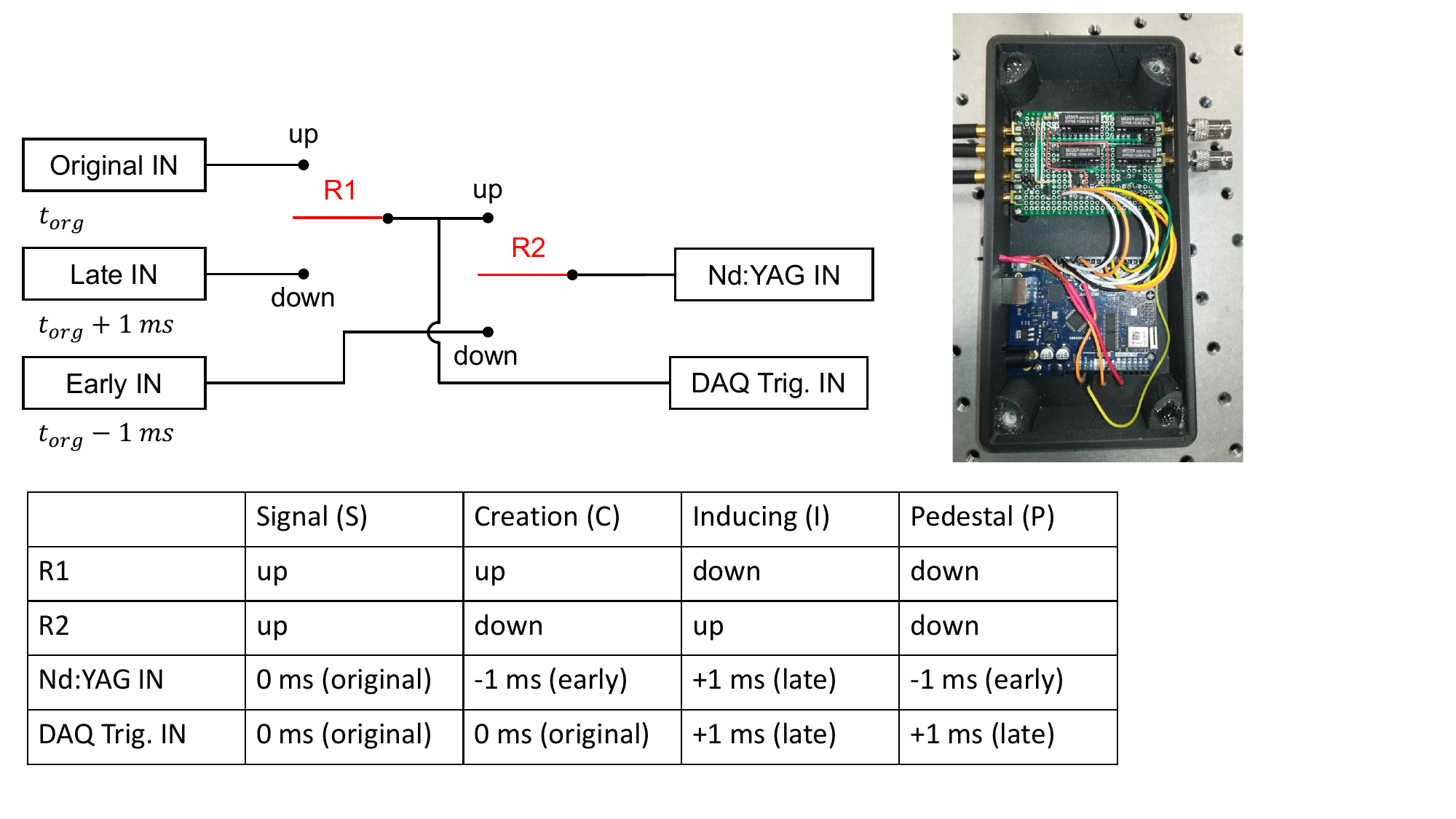}
\caption{\label{Fig_trigger_pattern_generator} 
Schematic diagram shows the wiring configuration in the relay circuit 
to control the Nd:YAG laser emission (Q-switch timing) and the start timing of data acquisition.
The table summarize the states of relay switches and the resulting timing shifts for the output trigger signal.
The photograph shows the relay circuit installed in the experimental area, with two additional relays intended for functionality that remained unused. 
}
\end{figure}

The TPG generates these patterns by shifting the Q-switch timing for the Nd:YAG laser operation and the start timing 
for the waveform acquisition via a simple relay circuit (Fig. \ref{Fig_trigger_pattern_generator}). 
To avoid residual effects created by long tailing component in time (e.g., spontaneous emission of Nd:YAG laser), 
a $\pm$ 1 ms timing separation is applied for the desynchronization of the laser pulses and the DAQ trigger timing. 
The $\pm$ 1 ms shifted TTL signals are generated from a local timing module (Fig. \ref{fig_trigger_system_setup}). 
The TPG utilizes mechanical relays to avoid introducing additional jitter by electronic components to route these signals, 
and switches the relays between laser shots. 

\begin{figure*}[t]
\centering
\includegraphics[scale=0.29]{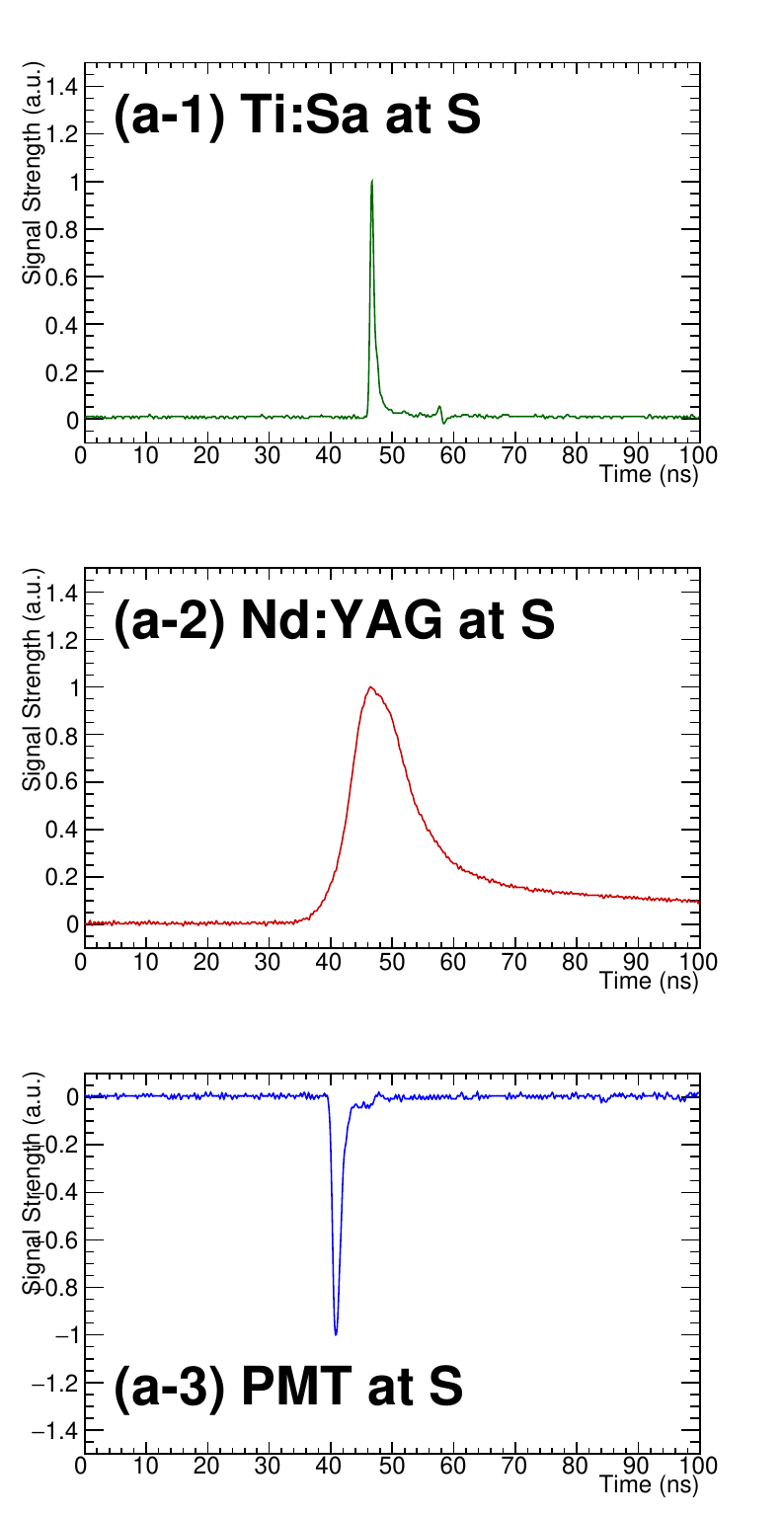}
\includegraphics[scale=0.29]{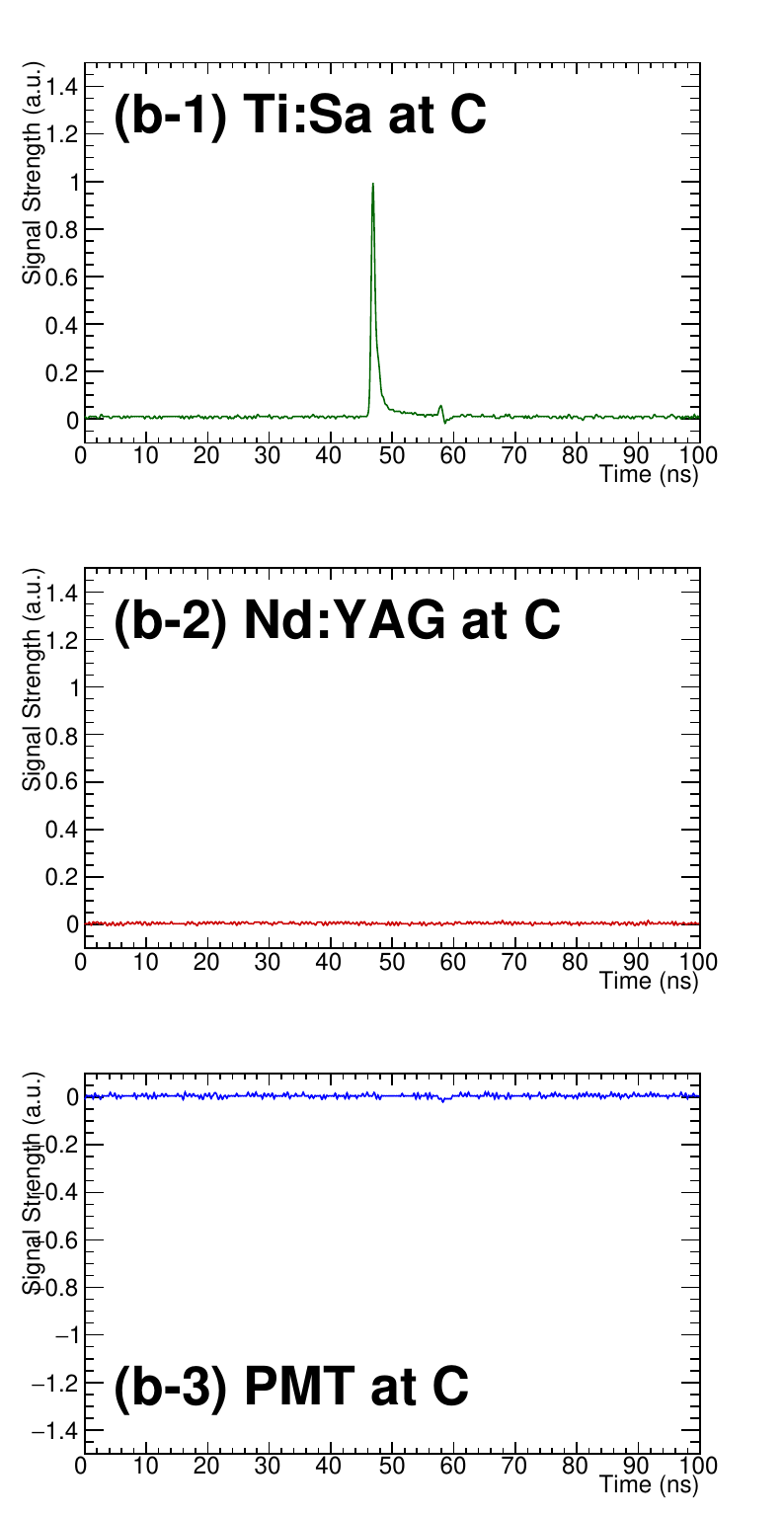}
\includegraphics[scale=0.29]{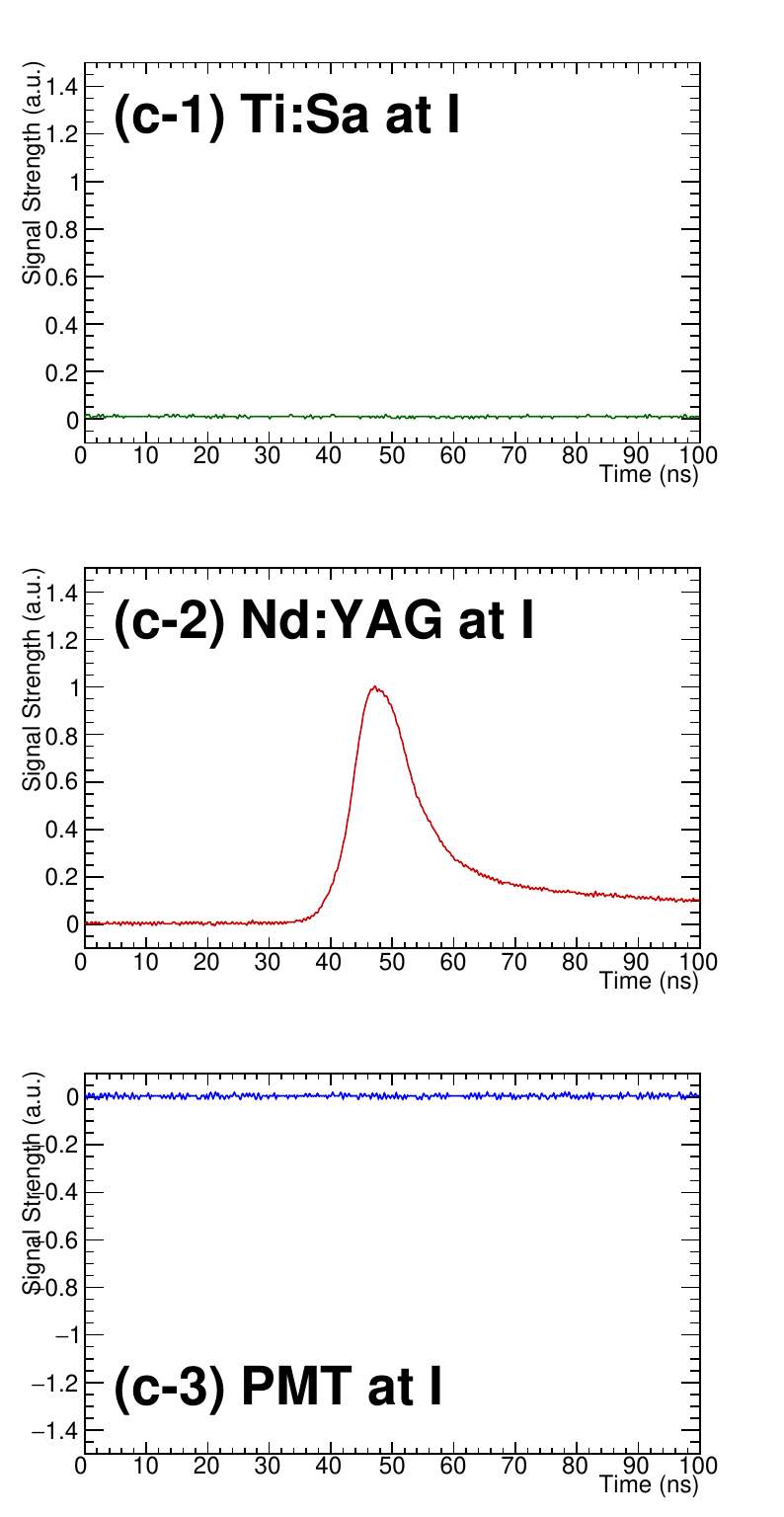}
\includegraphics[scale=0.29]{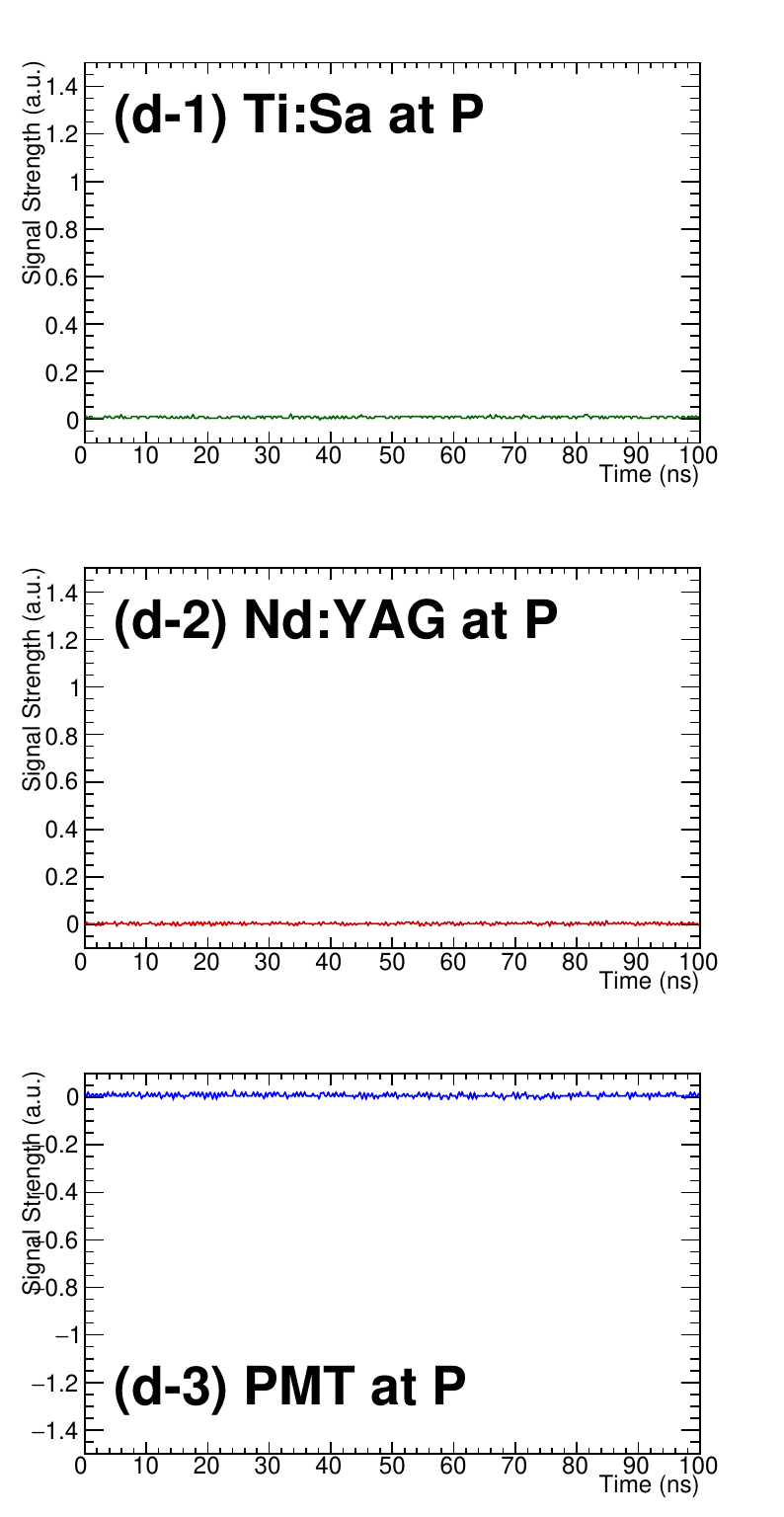}
\caption{\label{Fig_FWM_Air} Single-shot waveform in consecutive four laser shots 
at the atmospheric pressure with trigger pattern (a) "S", (b) "C", (c) "I", (d)) "P".
Top figures show the output from the photodiodes (PD1 in Fig. \ref{fig_opticalsetup_E4}) 
for Ti:Sa creation laser, 
the middle figures show the output from photodiodes (PD2 in Fig. \ref{fig_opticalsetup_E4}) 
for Nd:YAG inducing laser,
and the bottom figures show the output from photomultiplier tube (PMT in Fig. \ref{fig_opticalsetup_E4}) 
for the FWM signal detection. 
Each waveform is normalized at peak amplitude for visibility of the patterns. 
}
\end{figure*}

To ensure the functionality of the signal detection system, FWM signals were deliberately generated 
at the atmospheric pressure to enhance the signal strength. 
The detection system was verified in terms of the spatiotemporal overlap of the lasers 
and the trigger pattern control. Raw signals of the lasers are measured by the two photodiodes used 
for timing monitors. 
The FWM signal was measured by a PMT (R9800U-20, Hamamatsu) at the end of Signal Detection (Fig.\ref{fig_opticalsetup_E4}).
Figure \ref{Fig_FWM_Air} shows single-shot waveforms measured in the oscilloscope 
within a 100 ns time window via the start timing control of the DAQ trigger.
The spatial overlap at the focal point was maintained during the waveform measurements, 
only timing desynchronization was introduced by varying the trigger pattern.
The first three figures from the left show the result at the triggerr pattern "S" such 
that Ti:Sa (Fig.\ref{Fig_FWM_Air} (a-1)) and Nd:YAG laser (Fig.\ref{Fig_FWM_Air} (a-2)) 
were fired in timing synchronization. The FWM signal at the PMT (Fig.\ref{Fig_FWM_Air} (a-3)) was seen
 only at the "S" pattern.
The waveforms at the "C", "I", "P" pattern are shown in Fig.\ref{Fig_FWM_Air} (b), (c), and (d) respectively.
The presence and absence of the waveforms within 100 ns time window show consistentcy with
 the designed trigger pattern generation (Fig.\ref{Fig_trigger_pattern}).

\begin{figure}[H]
\centering
\includegraphics[scale=0.35]{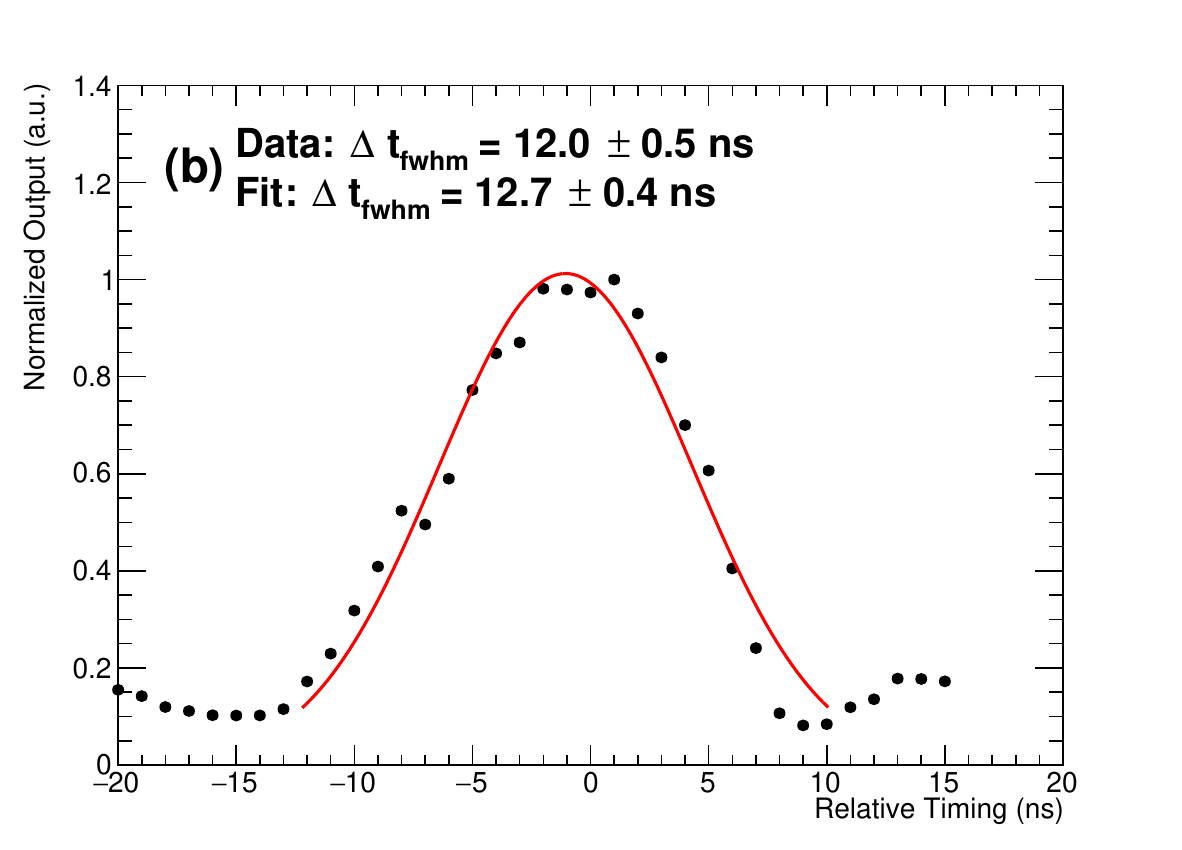}
\caption{\label{Fig_TimingSync} The FWM generation at atmospheric pressure as a function of Q-switch timing 
for the Nd:YAG laser operation.}
\end{figure}

A further cross-evaluation of the FWM signal detection and the timing synchronization of two lasers was performed 
via the timing dependence of the FWM genearation. Figure \ref{Fig_TimingSync} demonstrates that 
the FWM signal generation favors the temporal shape of Nd:YAG laser. 
The integrated charge of the raw waveform is plotted in Fig. \ref{Fig_TimingSync} (b) 
as a function of the relative timing of the Q-switch external trigger timing for the Nd:YAG laser operation. 
The pulse duration evaluated by the FWM generation shows comparable results with those 
from the direct measurement of Nd:YAG laser as shown in Fig. \ref{Fig_Temporal_NdYAG}.

\section{Commissioning measurements of four-wave mixing background photons in vacuum}

\begin{figure}[H]
 \centering
 \includegraphics[scale=0.3]{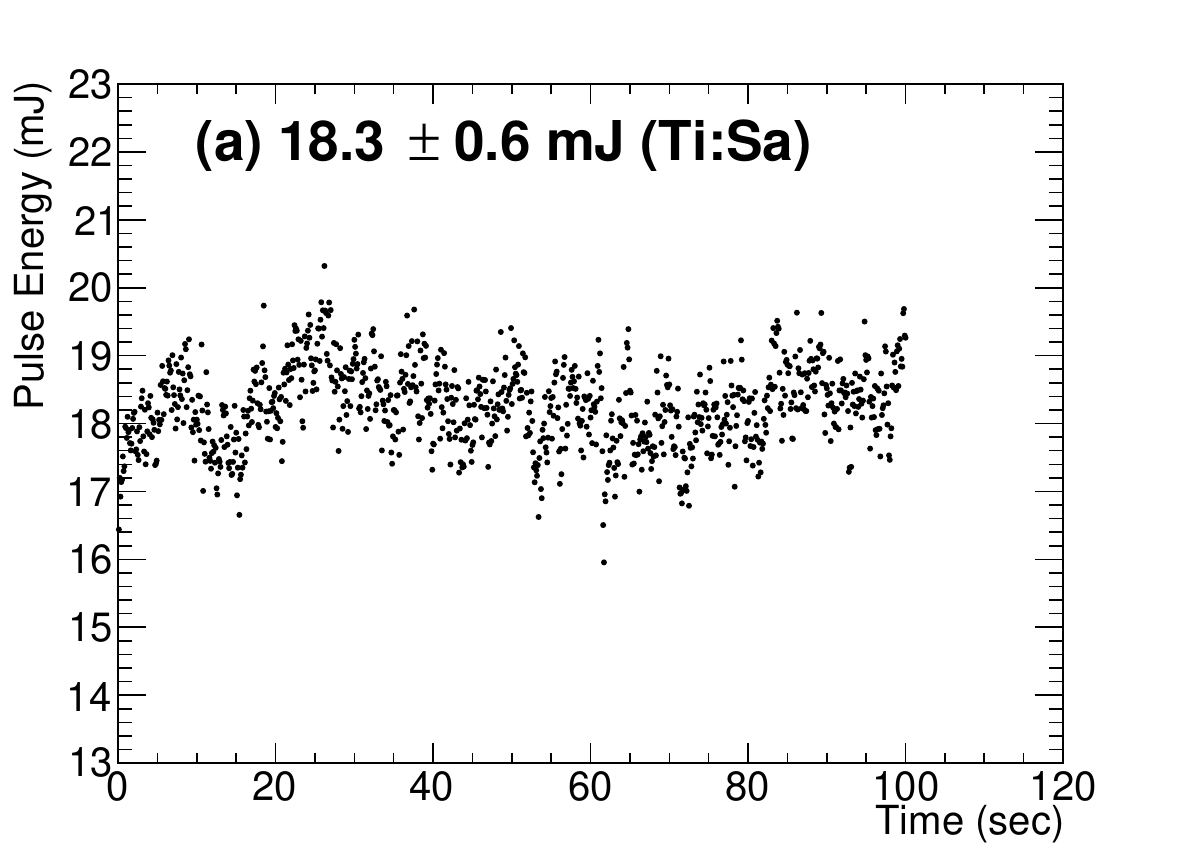}
 \includegraphics[scale=0.3]{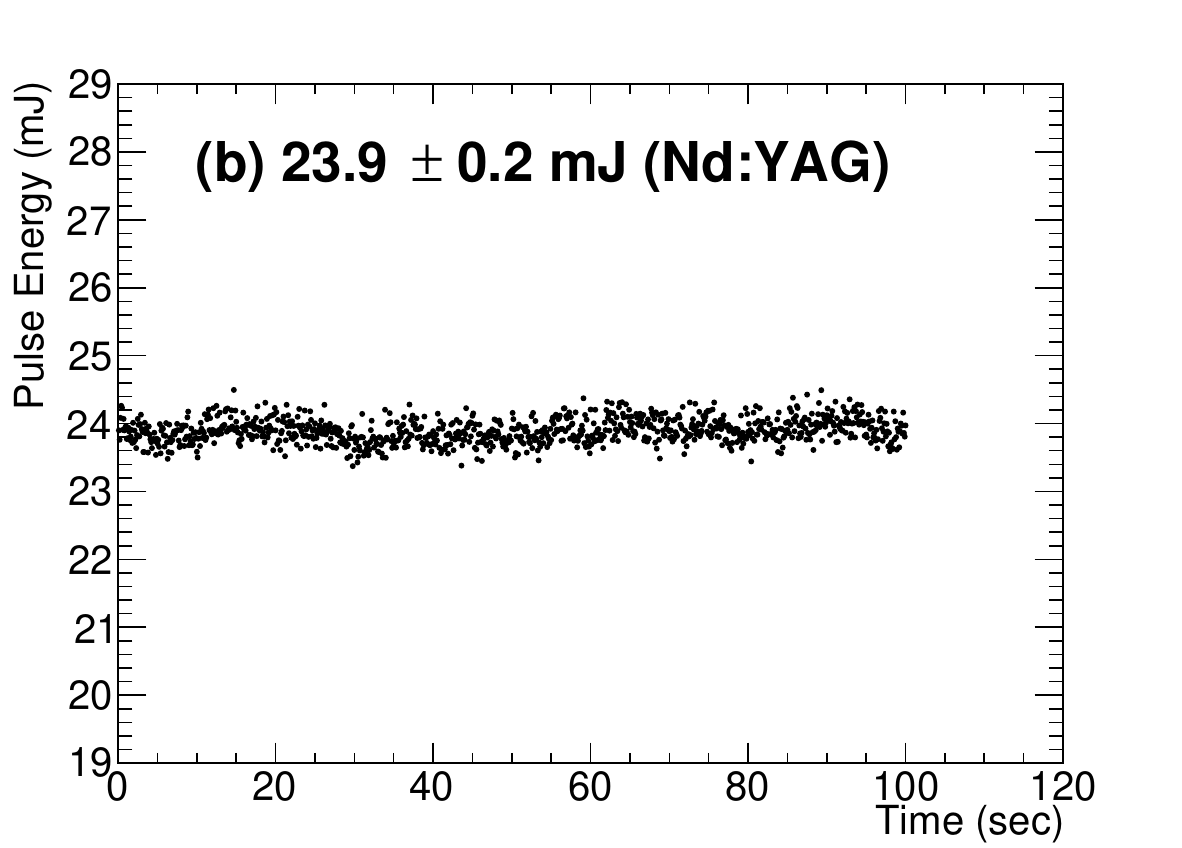}
 \caption{\label{fig_Pulse_Energy} Pulse energies for (a) Ti:Sa creation laser and (b) Nd:YAG inducing laser.
 Shot-by-shot energy flucutations over 1000 consecutive laser shots are shown.}
 \end{figure}

Commissioning measurements of the FWM signal were performed using the interaction of the two lasers 
with pulse energies of $18.3 \pm 0.6$ mJ for Ti:Sa creation laser and $23.9 \pm 0.2$ mJ 
for Nd:YAG inducing laser, respectively.
Figure \ref{fig_Pulse_Energy} shows the measured energies for Ti:Sa (left) and Nd:YAG (right) 
over 1000 laser shots.
We initiated these commissioning studies at energy levels similar to 
those used in our previous research \cite{JHEP2025_SAPPHIRES_10mJ}.

\begin{figure*}[t]
\centering
\includegraphics[scale=0.29]{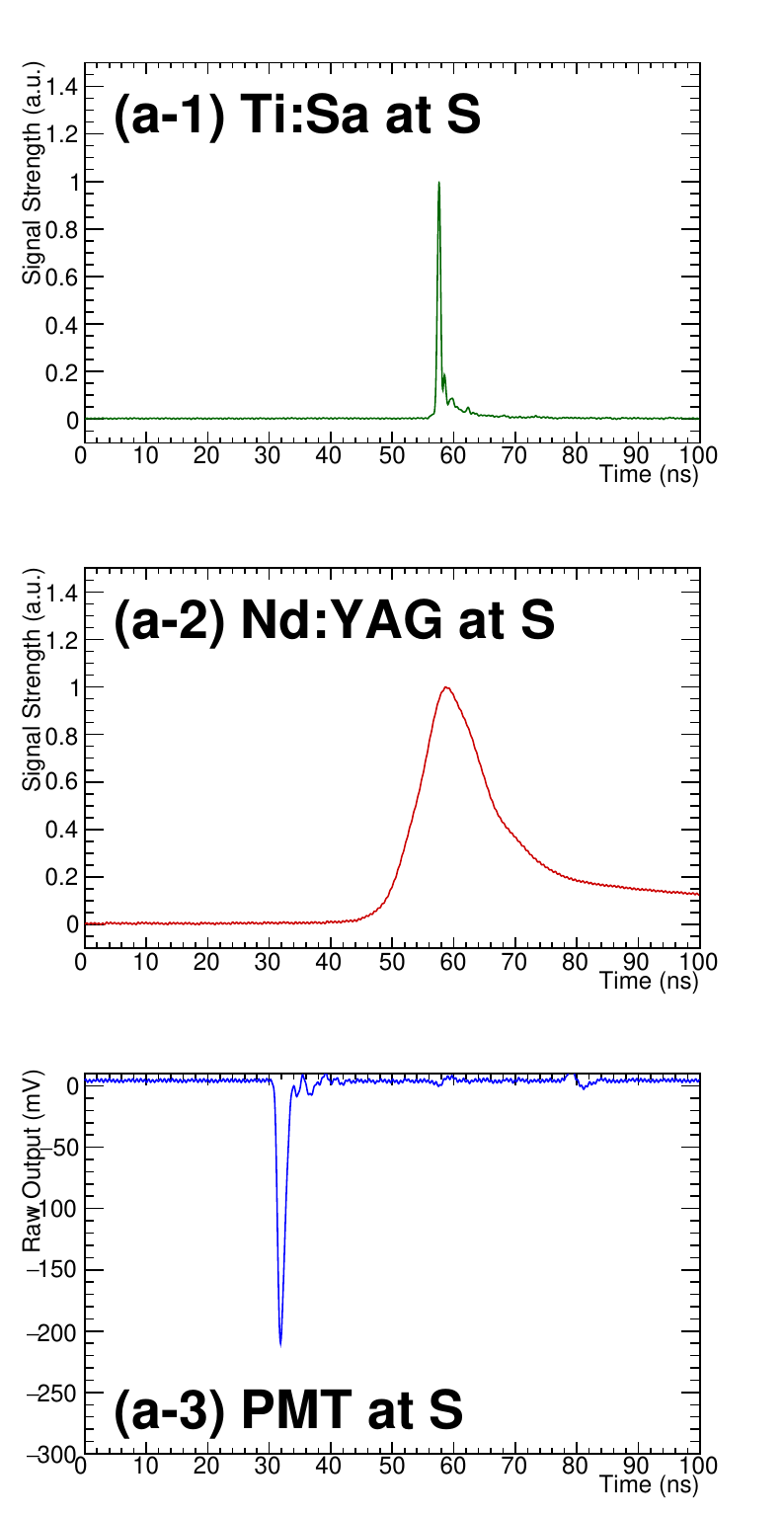}
\includegraphics[scale=0.29]{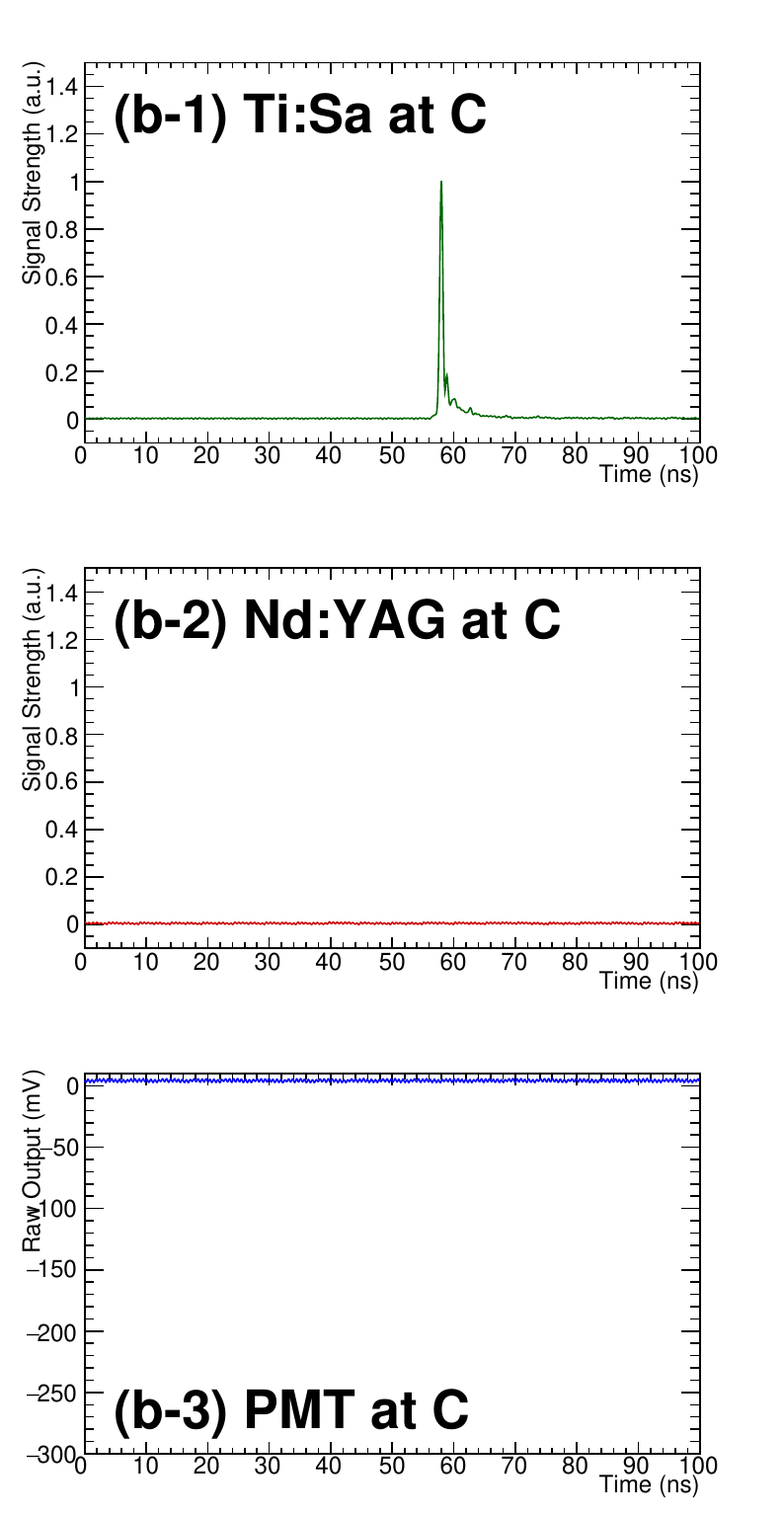}
\includegraphics[scale=0.29]{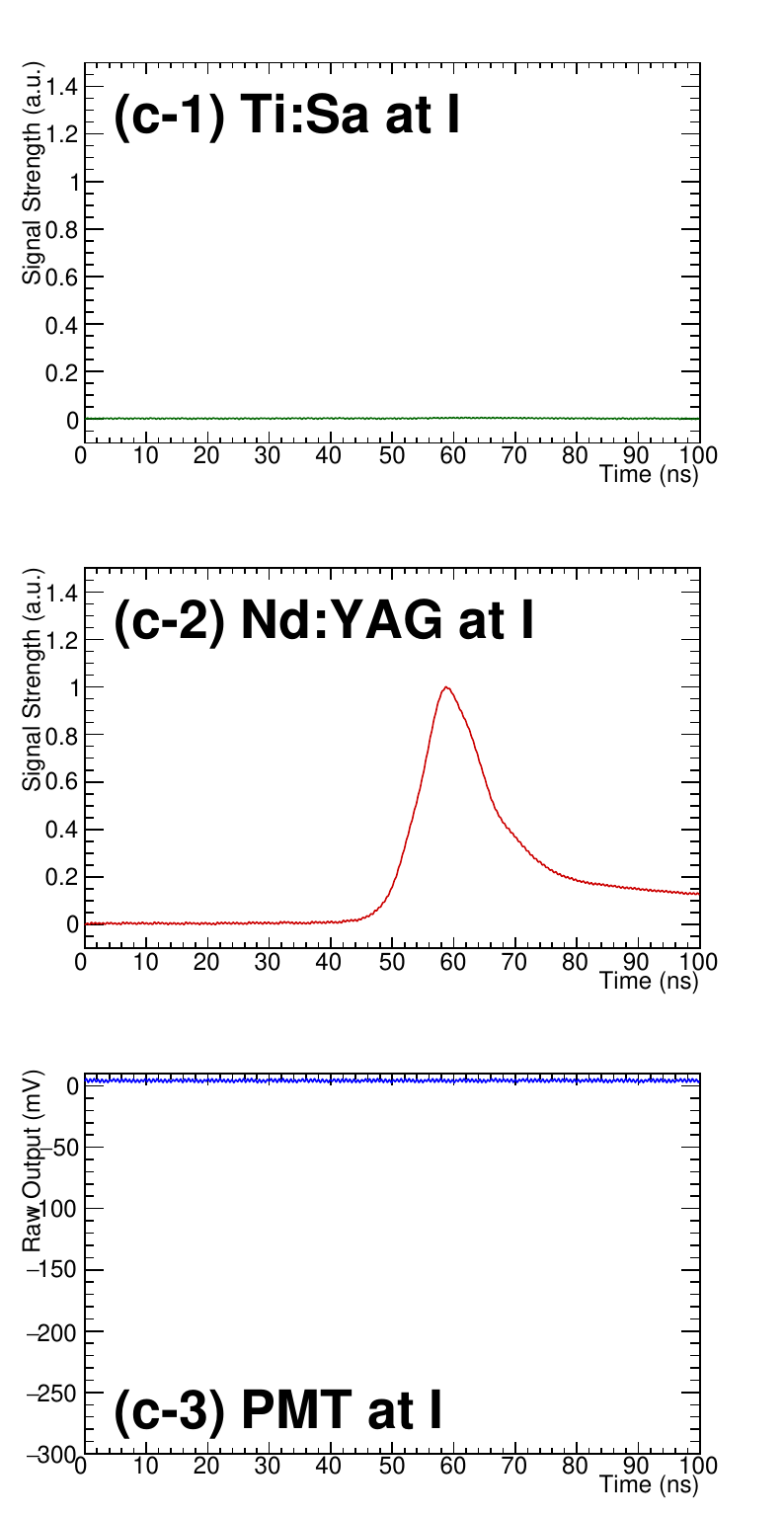}
\includegraphics[scale=0.29]{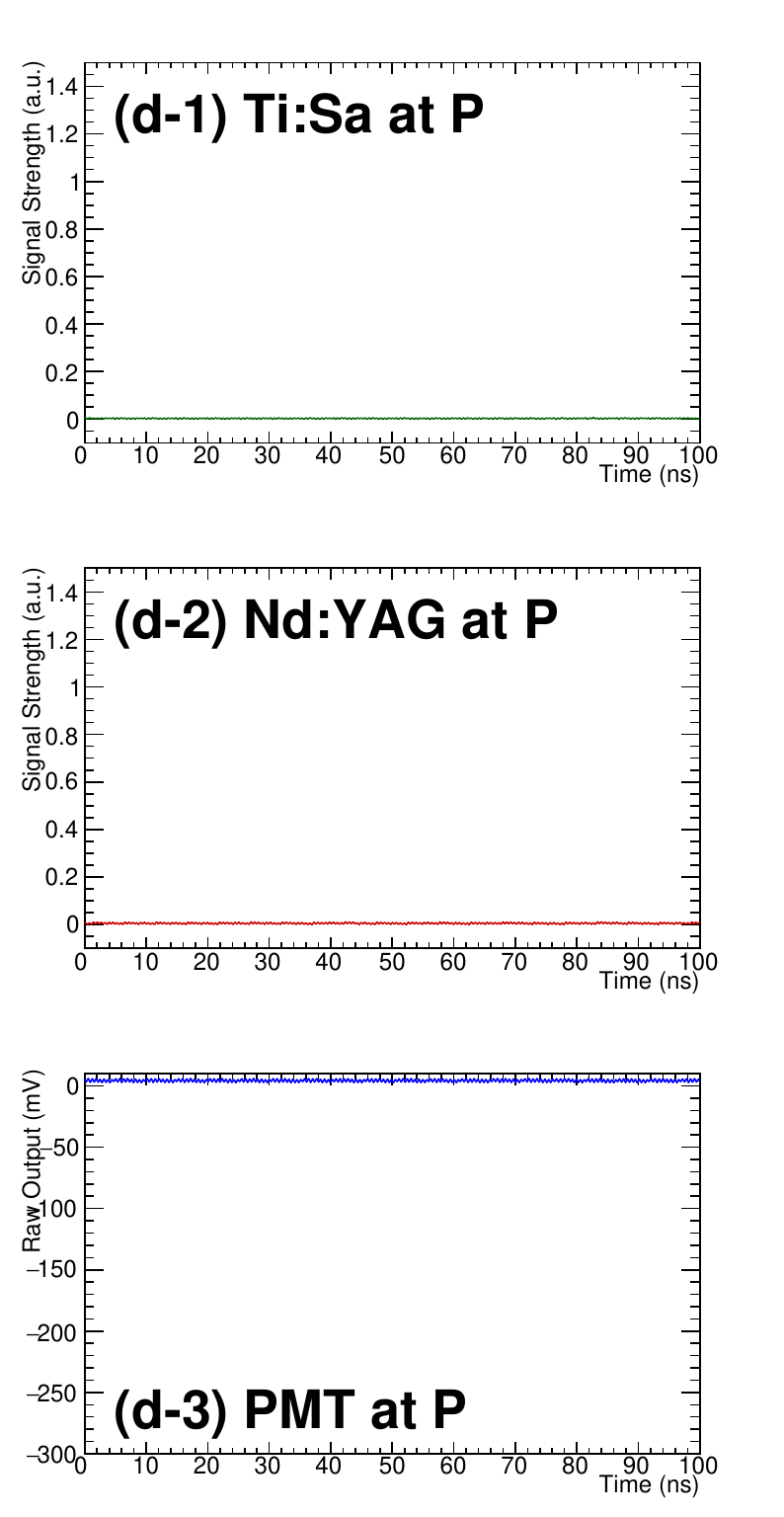}
\caption{\label{Fig_FWM_VAC}  1500-shot averged Waveforms for each trigger pattern (a) "S", (b) "C", (c) "I", and (d) "P". 
The top row displays the outputs from the photodiode (PD1 in Fig.\ref{fig_opticalsetup_E4}) 
used for the energy and timing monitoring of Ti:Sa creation laser. 
The middle row shows the outputs from the photodiode (PD2 in Fig.\ref{fig_opticalsetup_E4}) 
for the energy and timing monitoring of Nd:YAG inducing laser. 
The bottom row presents the outputs from the photomultiplier tube (PMT in Fig.\ref{fig_opticalsetup_E4}) 
for the four-wave mixing signal detection. 
Waveform data were acquired with laser pulse energies of 18.3 mJ (Ti:Sa) and 23.9 mJ (Nd:YAG) at a vacuum pressure of 1.3 $\times 10^{-7}$ mbar.}
\end{figure*}

\begin{figure}[h]
\centering
\includegraphics[scale=0.35]{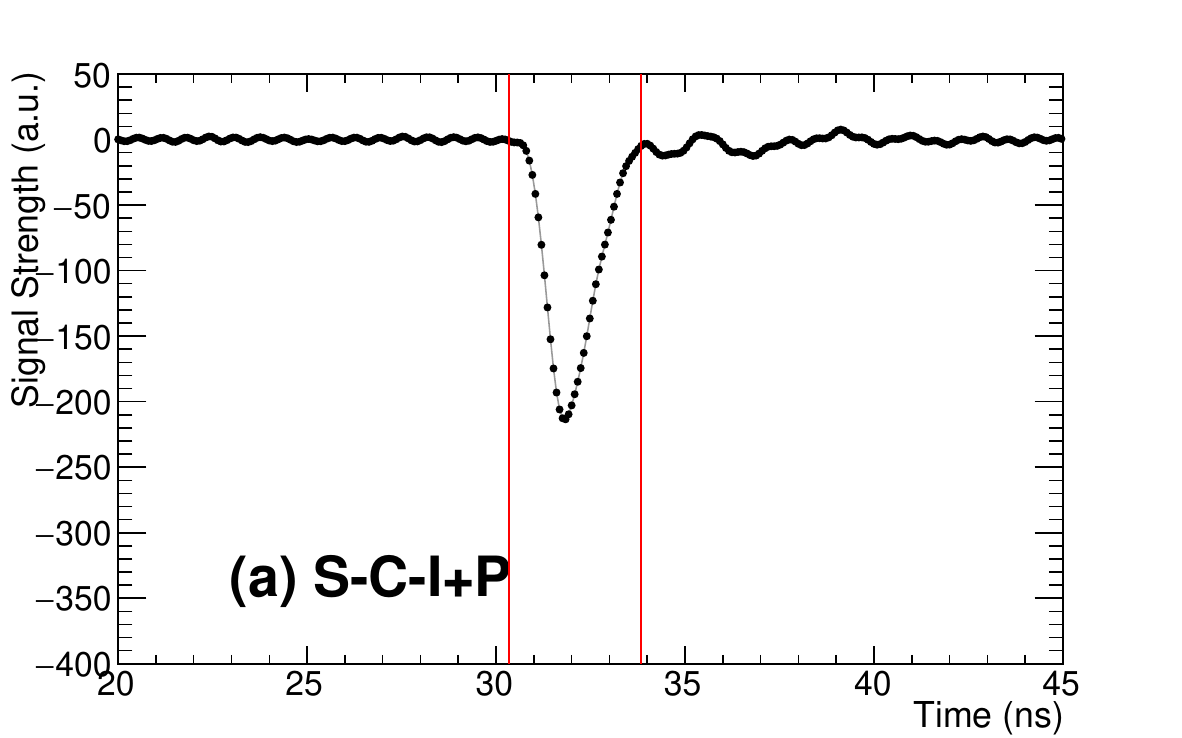}
\includegraphics[scale=0.35]{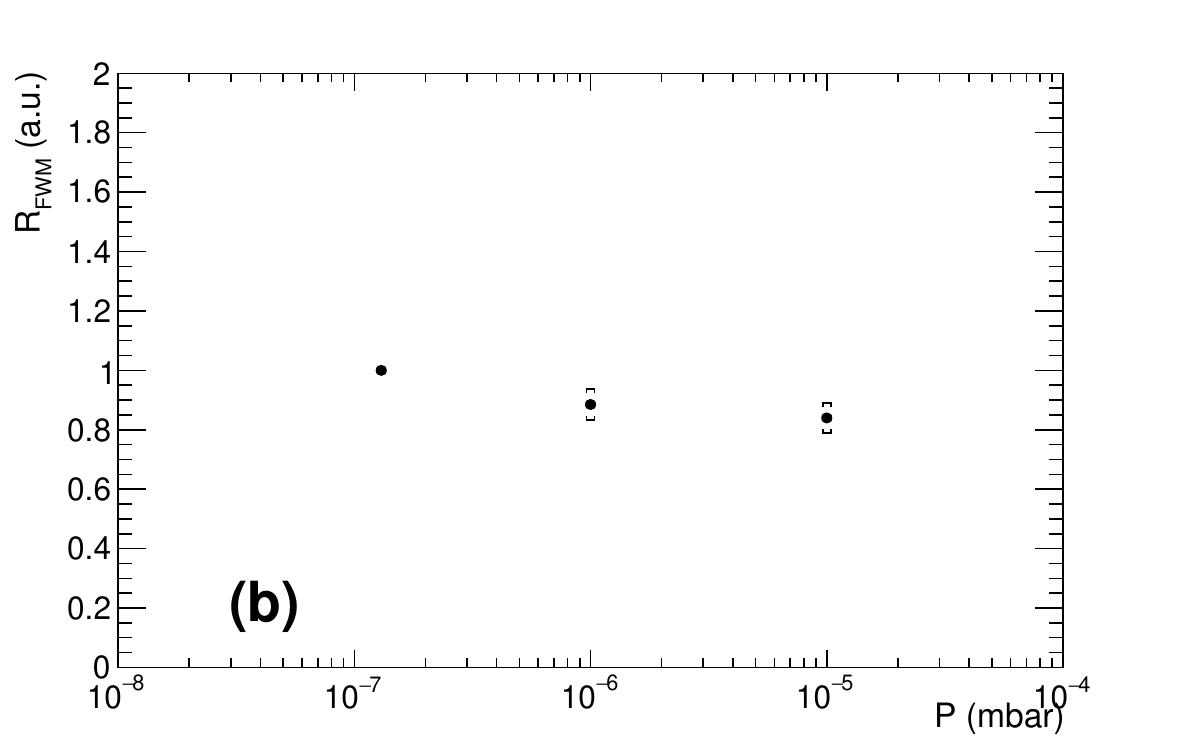}
\caption{\label{Fig_FWM_Pressure} (a) 1500-shot averaged net waveform at $1.3 \times 10^{-7}$ mbar
after subtracting the backgrounds based on the trigger events (S-C-I+P). 
(b) Pressure dependence of the ratio of integrated FWM signals based on 1500 laser-shot statistics. 
The integration of FWM signal was performed over the 3.5 ns time range depicted by the red vertical lines in (a).
The ratio $R_{FWM}$ is defined with respect to the reference data point at $1.3 \times 10^{-7}$ mbar. 
}
\end{figure}

The operational vacuum pressure was $1.3 \times 10^{-7}$ mbar, which represents the near-optimal pressure 
achievable in the VE1 vacuum system within a reasonable pumping time for a given beamtime. 
Figure \ref{Fig_FWM_Pressure} shows the waveforms to illustrate the signal patterns 
at different trigger classes: "S", "C", "I", and "P". 
The observed signal patterns are similar to the results at atmospheric pressure (Fig. \ref{Fig_FWM_Air}).
A conlusion from the results is that the background processes, such as those from the plasma ($n_{plasma}^{gas}$
and $n_{plasma}^{opt}$) and the defect level of the glass substrate $n_{FS}$, are suppressed to be negligibly 
small amounts at this vacuum pressure level because the FWM signal was observed with a significant signal strength 
only for the trigger pattern "S",  while those for the other trigger patterns were nearly null. 

To further identify the background sources, we investigated the pressure dependence of FWM generation. 
The net waveform of the FWM generation was obtained by dealing the raw waveforms as S - C - I + P.
Figure \ref{Fig_FWM_Pressure} (a) shows a typical net wavefrom of the FWM geneartion 
at $1.3 \times 10^{-7}$ mbar. The two vertical lines on the plot indicate the integration range 
of 3.5 ns around the FWM signal peak. 
Figure \ref{Fig_FWM_Pressure} (b) shows the ratio of the FWM generation 
in the pressure range from $1.3 \times 10^{-7}$ to $1.0 \times 10^{-5}$ mbar, 
normalized to the results at $1.3 \times 10^{-7}$ mbar. 
1500 laser-shot events are used for the analysis for each trigger pattern. 
The results suggest that the atomic FWM background photons $n_{aFWM}$ are also suppressed, 
and a power-law scaling, which is characteristic of the gas-origin background as shown in Fig.  \ref{fig_bg_species},
does not appear in that pressure range. 
Thus, at this energy level, the possible scenarios are either the ALP signal ($n_vFWM$) or, more likely, 
the optics-origin background ($n_{oFWM}$).
Therefore, the primary focus for the ALP searches at higher energies ( $>$ 0.1 J) should be 
the investigation of the optics-origin FWM backgrounds.

\section{Conclusion}
We have clarified the key upgrade issues for laser-driven axion-like particle (ALP) searches 
towards PW-class laser experiments, specifically addressing the categories of possible background processes 
relevant to the four-wave mixing (FWM) approach in a quasi-parallel photon-photon collision system. 
The implementation and commissioning of the experimental system were completed at the experimental area 
for 0.1 PW laser output at ELI-NP.
The experimental system is designed to minimize unwanted background and handle systematic uncertainties, 
and to characterize the pressure-dependent gas-origin backgrounds and the area-size dependent optics-origin backgrounds. 
Furthermore, a trigger pattern generator for the event-class selection was implemented 
to identify background sources for every four consecutive laser shots.

The performance of the experimental system was demonstrated. 
The vacuum pressure for gas-origin background measurements was controlled to below 1.3 $\%$ stability 
within the pressure range from $3 \times 10^{-7}$ to $3 \times 10^{-3}$ mbar. 
The functionality of the area-size control system was verified by measuring laser diameters 
using a knife-edge technique. The measured diameters before focusing were $59.7 \pm 1.1 $ mm 
for Ti:Sa creation laser and $19.9 \pm 1.6 $ mm for Nd:YAG laser.
The ratio of these diameters shows reasonable agreement with the inverse ratio of the focal-spot sizes obtained 
by independent measurements. 
The system facilitates the study of the scaling properties for the background generations. 

Spatiotemporal control of laser parameters is a critical factor not only 
for the laser-driven ALPs search but also 
for the commissioning of the new system at the experimental site. 
The focusing qualities of Ti:Sa high-power creation laser and Nd:YAG inducing laser were evaluated.
The spot diameters of Ti:Sa laser are $8.8 \, \pm 1.1 \, \mu$m (horizontal) and $7.6 \, \pm 0.5 \, \mu$m (vertical) 
at FWHM. The results are reasonably close to a diffraction-limited ideal diameter of 7.4 $\mu$ m calculated at 800 nm.
The spot diameters of Nd:YAG laser were $31.7 \, \pm 2.0 \, \mu$m (horizontal) and $27.4 \, \pm 3.2 \, \mu$m (vertical).
The results are consistent with the expected values from our optical design 
to ensure stable four-wave mixing generation 
against the unavoidable finite pointing instability.  
Pointing stability was measured over 49$\,$670 laser-shot events. 
The spatial overlap at peak intensities is guaranteed within a $\pm \, 0.5 \, \sigma$ range 
relative to Nd:YAG laser spot diameter. 
The temporal shapes of the lasers were also measured. 
The pulse duration of Ti:Sa laser is $26.1 \pm 0.7 $ fs which is consistent with the Fourier-transform limit 
calculated by independent spectral measurements for Ti:Sa laser. 
Regarding Nd:YAG laser, the pulse duration of  $10.5 \pm 0.3$ ns was obserbed. 
Timing synchronization between the lasers is maintained by a low-jitter timing system, 
achieving a total timing jitter of 0.35 ns as standard deviation. 
The results show that timing synchronization is sufficiently controlled to stay within a 
$\pm 0.5 \, \sigma$ equivalent timing shift relative to the temporal peak of Nd:YAG laser pulse. 

The waveform detections for time-resolved ALP signal observation were validated 
under enhanced atomic FWM generation conditions at atmospheric pressure. 
Expected signal waveforms were obtained across consecutive laser-shot events with different trigger patterns. 
Furthermore, the signal detection capability was demonstrated by measuring Nd:YAG laser pulse duration 
via atomic FWM process. We conclude that the key features for the FWM detections are fully functional 
for identifying and understanding possible background sources towards the 0.1 PW laser experiment at ELI-NP.
 
The strategies for FWM background studies were validated with integrated experimental system 
developed at ELI-NP. This was performed through the waveform analysis at a vacuum pressure of $1.3 \times 10^{-7}$ mbar
and its pressure dependence at 20 mJ-level laser pulse energies,
 specifically 18.3 mJ for Ti:Sa creation laser and 23.9 mJ for Nd:YAG inducing laser, respectively.
A power-law scaling of vacuum pressure $P^{\alpha} \left(\alpha = 2 - 3 \right)$ was not observed for the FWM generation in the pressure range of $1.3 \times 10^{-7}$ to $1.0 \times 10^{-5}$ mbar. Such 
pressure insensitive FWM generation, likely originating from the optics-origin background photons, 
poses an important challenge for ALP searches towards 2.5J full energy in 0.1 PW laser system. 
We conclude that the implemented experimental system provides a suitable platform 
for the background studies and enables a stepwise scale-up of the laser pulse energy 
to 0.1 J, 1J, and 2.5 J for ALPs search at ELI-NP.

\section*{Acknowledgment}
This work was supported by contracts PN23210105 and ELI-RO/DEZ/2023 $\_$001 
funded by the Romanian Ministry of Education and Scientific Research, 
and the Extreme Light Infrastructure Nuclear Physics Phase II project, 
co-financed by the Romanian Government and the European Union through the European Regional Development Fund 
and the Competitiveness Operational Programme (No. 1/07.07.2016, COP, ID 1334). 
The ELI-NP user facility operation is supported by the Romanian Ministry of Education 
and Scientific Research through the National Interest Infrastructure Facility IOSIN-ELI.

K. Homma acknowledges the support of the Collaborative Research Program of the Institute for Chemical Research at Kyoto University (Grant Nos. 2024-95
and 2025-100), the JSPS Core-to-Core Program (grant number: JPJSCCA20230003), and Grants-in-Aid for 
Scientific Research (Nos. and 24KK0068 and 25H00645)
from the Ministry of Education, Culture, Sports, Science and Technology (MEXT) of Japan. 

K. A. Tanaka acknowledges the support by JSPS Core-to-Core Program, (grant number: JPJSCCA20230003).

We are grateful to Laser System Department and Technical Department at ELI-NP for their technical support
during the commissioning of the experiment.

\vspace{0.2cm}
\noindent

\let\doi\relax

\end{document}